\documentclass[prd,twocolumn,showpacs,nofootinbib,preprintnumbers]{revtex4}

\usepackage{pictex}
\usepackage{epsfig}
\usepackage{amsmath}
\usepackage{amssymb}

\def\e{{\,\rm e}\,}
\def\d{{\rm d}}
\def\i{{\rm i}}

\def\N{{N}}
\def\Nf{{N_f}}
\def\A{{\cal A}}

\def\F{{\cal F}}
\def\T{{T}}
\def\Beta{{\cal B}}
\newcommand{\tr}[1]{\,{\rm tr}\,#1}

\newcommand{\rf}[1]{(\ref{#1})}
\newcommand{\eq}[1]{Eq.~(\ref{#1})}
\def\be{\begin{equation}}
\def\ee{\end{equation}}
\def\beq{\begin{equation}}
\def\eeq{\end{equation}}
\def\bea{\begin{eqnarray}}
\def\eea{\end{eqnarray}}
\def\LA{\left\langle}
\def\RA{\right\rangle}
\def\D{{\cal D}}

\newcommand{\non}{\nonumber \\*}
\newcommand{\ie}{{i.e.}\ }

\font\thinlinefont=cmr5
\begingroup\makeatletter\ifx\SetFigFont\undefined
\def\x#1#2#3#4#5#6#7\relax{\def\x{#1#2#3#4#5#6}}%
\expandafter\x\fmtname xxxxxx\relax \def\y{splain}%
\ifx\x\y   % LaTeX or SliTeX?
\gdef\SetFigFont#1#2#3{%
  \ifnum #1<17\tiny\else \ifnum #1<20\small\else
  \ifnum #1<24\normalsize\else \ifnum #1<29\large\else
  \ifnum #1<34\Large\else \ifnum #1<41\LARGE\else
     \huge\fi\fi\fi\fi\fi\fi
  \csname #3\endcsname}%
\else
\gdef\SetFigFont#1#2#3{\begingroup
  \count@#1\relax \ifnum 25<\count@\count@25\fi
  \def\x{\endgroup\@setsize\SetFigFont{#2pt}}%
  \expandafter\x
    \csname \romannumeral\the\count@ pt\expandafter\endcsname
    \csname @\romannumeral\the\count@ pt\endcsname
  \csname #3\endcsname}%
\fi
\fi\endgroup

\begin{document}

\preprint{ITEP--TH--21/09}

\title{A Brief Introduction to Wilson Loops and Large $N$}

\author{Yuri Makeenko}
%\altaffiliation[Also at]{ the Institute for Advanced Cycling,
%Blegdamsvej 19, 2100 Copenhagen \O, Denmark}

\affiliation{Institute of Theoretical and Experimental Physics,
 Moscow, Russia}
\email{makeenko@itep.ru}

%%\date{\mbox{}}

\begin{abstract}
A pedagogical introduction to Wilson loops, lattice gauge theory
and the $1/N$-expansion of QCD is presented.
The three introductory lectures were given at the 37th ITEP Winter
School of Physics, Moscow, February 9--16, 2009.

\end{abstract}


\maketitle

\section{Introduction}

In these lecture notes I give a brief pedagogical introduction to
the methods used in nonperturbative investigations of QCD and 
other gauge theories. The main attention is payed to Wilson loops,
both on the lattice and in the continuum, which play a central role
in modern formulations of gauge theories and to the method of the
$1/N$ expansion.

For further studies of this subject I can recommend the 
textbook~\cite{Mak02} which contains detail references.
In this text I restrict myself only with references to a few classical
papers.

\part{Wilson loops}

Wilson loops are essentially phase factors in 
Abelian or non-Abelian gauge theories.
Wilson loops are observable in quantum theory by the {Aharonov--Bohm} 
effect.
Wilson loops play a central role in the {lattice formulation} of gauge 
theories.
QCD can be reformulated through the Wilson loops in a manifest gauge-invariant
way.
Analogues of the Wilson loops are extremely useful in solving 
various kinds of matrix models.

\section{Phase factors in QED}

\subsection{Definition and basic properties}

Abelian {phase factor}\index{phase factor!Abelian} is defined by
the formula
\bea
U[\Gamma_{yx}]&=& \e^{ \i e \int_{\Gamma_{yx}} \d z^\mu A_\mu(z)}.
\label{aphasefactor}
\eea
Under the {gauge transformation}\index{transformation!gauge}
\begin{eqnarray}
A_\mu ( z) &
\stackrel{{\rm g.t.}}{\longrightarrow}&
 A_\mu ( z) +\frac
1{e}\partial _\mu \alpha ( z)\,,
\label{abgaugetr}
\end{eqnarray}
the Abelian phase factor\index{phase factor!Abelian} {transforms} as
\begin{eqnarray}
U[{\Gamma_{yx}}] & \stackrel{{\rm g.t.}}{\longrightarrow }&
\e^{\i\alpha ( y) } \,
U[{\Gamma_{yx}}]
\e^{-\i\alpha ( x) } .
\label{86}
\end{eqnarray}

A {wave function} at the point $x$ is {transformed}
%%under the gauge transformation~\rf{abgaugetr} 
as\index{transformation!gauge!matter field}
 \bea
 \varphi(x)&
\stackrel{{\rm g.t.}}{\longrightarrow }&
 \e^{\i\alpha(x)}\, \varphi(x)\,,
 \label{transformation}
 \eea
therefore the phase factor\index{phase factor!Abelian}
is {transformed} as the product
$\varphi(y) \varphi^\dagger(x)$:
\bea
U[{\Gamma_{yx}}] &
\stackrel{{\rm g.t.}}{\sim}&
`` \varphi(y) \,\varphi^\dagger(x)\hbox{''}.
\eea

A {wave function} at the point $x$
transforms like one at the point $y$ after multiplication by
the phase factor\index{phase factor!Abelian}:
\bea
U[{\Gamma_{yx}}] \,\varphi(x) &
\stackrel{{\rm g.t.}}{\sim}&
`` \varphi(y)\hbox{''} ,
\label{asx}
\eea
and analogously
\bea
\varphi^\dagger(y)\, U[{\Gamma_{yx}}]  &
\stackrel{{\rm g.t.}}{\sim}&
`` \varphi^\dagger(x) \hbox{''} .
\label{asy}
\eea

The phase factor\index{phase factor} plays the role of a
{parallel transporter\index{parallel transport}} 
in an electromagnetic field, and  to compare phases 
of a wave function at points $x$ and $y$,
we should first make a parallel transport along some contour $\Gamma_{yx}$.
The result is {$\Gamma$-dependent} except when $A_\mu(z)$
is a pure gauge (vanishing field strength 
$F_{\mu\nu}(z)$). 
Certain subtleties occur for {not simply connected spaces}
(the {Aharonov--Bohm} effect).

\subsection{Propagators\index{propagator!in external field}
                       in external field \label{e.m.}}

Let us consider a {quantum} particle in a {classical} electromagnetic field. 
To introduce {electromagnetic field}, $\partial_\mu$ is to be replaced by the 
{covariant derivative}
\be
\partial _\mu  \longrightarrow  \nabla _\mu ~=~\partial _\mu 
-\i eA_\mu\! \left(x\right). 
\ee

For the propagator we get
\begin{eqnarray}
%\hspace*{2.5mm}
\lefteqn{%\hspace*{-5mm}
G\!\left( x,y;A\right)
=  \frac 12\int\nolimits_0^\infty \d\tau \e^{-\frac 12\tau m^2}  }\non &&\times
\!\!\!\!\! \int\limits_{\scriptstyle
z_\mu( 0) =x_\mu \atop \scriptstyle
z_\mu( \tau ) =y_\mu}\!\!\!\!\!  
\D z_\mu( t) \e^{-\frac 12\int_0^\tau \d t\,\dot z^2_\mu(t) +
\i e\int_0^\tau \d t\,\dot z^\mu( t) 
A_\mu(z( t)) } .  
\nonumber\\[-6mm]
& & 
\label{emfina}
\end{eqnarray}
The exponent is just the classical 
({Euclidean}) {\em action}\/\index{classical action}  of a particle in
an external electromagnetic field.
The path-integral representation~\rf{emfina} 
for the propagator\index{propagator!in external field} of a scalar particle 
in an external electromagnetic field is due to {Feynman}.

We can alternatively rewrite \eq{emfina} as
\begin{eqnarray}
G( x,y;A) & = &\sum\limits_{\Gamma_{yx}}
\e^{-S_{\rm free}[ \Gamma_{yx} ] +\i e \int_{\Gamma_{yx}} \d z^\mu A_\mu(z) } ,
\label{emsumovergamma}
\end{eqnarray}
where we represented the ({parametric invariant})\index{symmetry!parametric} 
integral over $\d t$ as the
contour integral along the {trajectory} $\Gamma_{yx}$ over
\bea
\d z^\mu &=& \d t \, \dot{z}^\mu (t)\,.
\label{dz=dtdotz}
\eea

The {transition amplitude} of
 a {quantum particle} in a {classical electromagnetic field}
is the sum over paths of the Abelian 
{phase factor}~\rf{aphasefactor}.

\subsection{Aharonov--Bohm effect}

{Transverse components}  of the electromagnetic field
describe photons. {Longitudinal} 
components
are related to gauging the phase of a wave function, \ie permit one
to {compare} its values at {different space-time points}
 when an electron is 
placed in an external electromagnetic field. 

In quantum 
mechanics, the wave-function phase itself is unobservable. Only the phase 
differences are observable, {\it e.g.}\/\ via interference phenomena.
 The phase difference 
depends on the value of the phase factor for a given path 
$\Gamma_{yx}$ along which the parallel transport is performed.

\begin{figure}
%\includegraphics{arbo}% Here is how to import EPS art
%\begin{figure}[tb]
%\unitlength=1.00mm
%%\unitlength=.80mm
\unitlength=.70mm
\linethickness{0.6pt}
\thicklines
%%\hspace*{-4cm}
\hspace*{-3.5cm}
\begin{picture}(105.00,90.00)(-10,-10)
%\begin{picture}(105.00,80.00)(-10,-4)
%%\begin{picture}(105.00,76.00)(5,0)
\put(50.00,0.00){\line(0,1){17.00}}
\put(50.20,0.00){\line(0,1){16.80}}
\put(49.90,0.00){\line(0,1){16.90}}
\put(50.10,0.00){\line(0,1){16.90}}
\put(49.80,0.00){\line(0,1){16.80}}
\put(50.00,19.00){\line(0,1){36.00}}
\put(50.10,19.10){\line(0,1){35.80}}
\put(49.90,19.10){\line(0,1){35.80}}
\put(50.20,19.20){\line(0,1){35.60}}
\put(49.80,19.20){\line(0,1){35.60}}
\put(50.10,57.10){\line(0,1){16.90}}
\put(49.90,57.10){\line(0,1){16.90}}
\put(50.20,57.20){\line(0,1){16.80}}
\put(49.80,57.20){\line(0,1){16.80}}
\put(50.00,57.00){\line(0,1){17.00}}
\put(60.00,37.10){\circle{10.00}}
\put(60.00,36.90){\circle{10.00}}
\put(60.10,37.00){\circle{10.00}}
\put(59.90,37.00){\circle{10.00}}
\put(60.00,37.20){\circle{10.00}}
\put(60.00,36.80){\circle{10.00}}
\put(60.00,37.10){\circle{10.00}}
\put(60.00,36.90){\circle{10.00}}
\put(60.20,37.00){\circle{10.00}}
\put(59.80,37.00){\circle{10.00}}
\put(60.10,37.00){\circle{10.00}}
\put(59.90,37.00){\circle{10.00}}
\put(60.00,37.00){\circle{10.00}}
\put(60.00,37.00){\makebox(0,0)[cc]{{\footnotesize $H\!\!\neq\!0$}}}
\put(12.00,37.00){\circle*{2.00}}
\put(16.00,37.00){\makebox(0,0)[lc]{{source}}}
%\put(16.00,37.00){\makebox(0,0)[lc]{source}}
\put(12.00,37.00){\vector(2,1){20.00}}
\put(12.00,37.00){\vector(2,-1){20.00}}
\put(32.00,47.00){\line(2,1){18.00}}
\put(32.00,27.00){\line(2,-1){18.00}}
\put(50.00,56.00){\vector(4,-1){24.00}}
\put(74.00,50.00){\line(4,-1){26.50}}
\put(50.00,18.00){\vector(2,1){24.00}}
\put(74.00,30.00){\line(2,1){26.50}}
\put(100.50,0.00){\line(0,1){74.00}}
\put(54.00,43.50){\makebox(0,0)[lb]{{solenoid}}}
\put(102.50,46.50){\makebox(0,0)[lc]{{interference}}}
\put(102.50,41.00){\makebox(0,0)[lc]{{plane}}}
\put(52.00,3.00){\makebox(0,0)[lc]{{screen}}}
\put(45.00,57.00){\makebox(0,0)[rc]{{electron}}}
\put(56.00,57.00){\makebox(0,0)[lc]{{beam}}}
\put(45.00,18.50){\makebox(0,0)[rc]{{electron}}}
\put(56.00,18.50){\makebox(0,0)[lc]{{beam}}}
%\put(54.00,43.00){\makebox(0,0)[lb]{solenoid}}
%\put(102.50,45.50){\makebox(0,0)[lc]{interference}}
%\put(102.50,41.00){\makebox(0,0)[lc]{plane}}
%\put(52.00,3.00){\makebox(0,0)[lc]{screen}}
%\put(45.00,57.00){\makebox(0,0)[rc]{electron}}
%\put(56.00,57.00){\makebox(0,0)[lc]{beam}}
%\put(45.00,18.50){\makebox(0,0)[rc]{electron}}
%\put(56.00,18.50){\makebox(0,0)[lc]{beam}}
\end{picture}
\caption{\label{fig:arbo} Scheme of the Aharonov--Bohm experiment.
}
\end{figure}
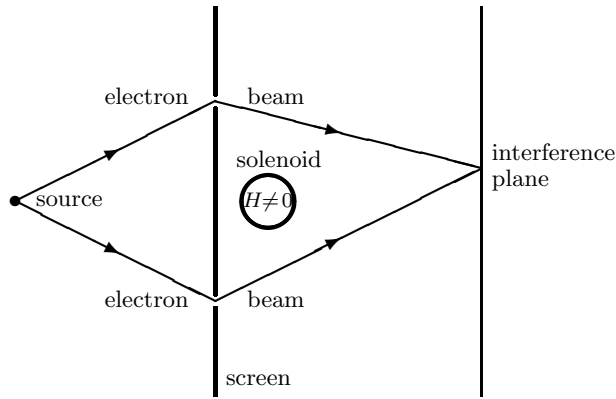

The phase factors are observable in quantum theory, 
in contrast to classical theory. This is seen in the {Aharonov--Bohm} 
effect~\cite{AB59} whose scheme is depicted in Fig.~\ref{fig:arbo}.
Electrons do not pass inside the solenoid
        where the magnetic field is concentrated. 
    Nevertheless, a phase difference arises between the electron beams passing 
through the two slits. 
The interference picture changes with the value of the electric current.
 
The phase difference depends on (the real part of)
\bea
%%{\rm Re} 
\lefteqn{\e^{\i e\int_{\Gamma^+_{yx}}\d z^\mu A_\mu ( z) }
\e^{-\i e\int_{\Gamma^-_{yx}}\d z^\mu A_\mu ( z) }}\non
&=&
%%{\rm Re}
\e^{\i e\oint_{\Gamma}\d z^\mu A_\mu ( z)} =  
\e^{\i e\int \d\sigma ^{\mu \nu }
 F _{\mu \nu }}   ~ = ~ \e^{\i eHS},
\eea
where the closed contour $\Gamma$ is composed from $\Gamma^+_{yx}$ and 
$\Gamma^-_{xy}$. 
It does not depend on the shape of
$\Gamma^+_{yx}$ and $\Gamma^-_{yx}$
but depends only on  $HS$ --- the magnetic flux through the solenoid.

\section{Yang-Mills theories}

Modern theories of fundamental interactions are {gauge theories}.
The principle of {local\index{symmetry!gauge} 
gauge invariance} was introduced by {H.~Weyl} for the
electromagnetic interaction in analogy with general covariance
in {Einstein}'s theory of gravitation. An extension to non-Abelian 
gauge groups was given by {Yang and Mills} in 1954~\cite{YM54}. 

A crucial role in gauge theories is played by 
the phase factor
which is associated with parallel transport\index{parallel transport} 
in an external gauge field. 
The phase factors are observable in quantum theory,
in contrast to classical theory. 
This is analogous to the {Aharonov--Bohm} effect for the electromagnetic field.

\subsection{Gauge invariance}

The principle of {local gauge invariance}\index{symmetry!gauge} 
deals with 
the {gauge transformation} (g.t.) of a matter 
field\index{transformation!gauge!matter field} 
$\psi$, which is given by 
\begin{eqnarray}
\psi ( x) & 
\stackrel{{\rm g.t.}}{\longrightarrow}
&\psi^\prime ( x) ~=~
\Omega\! \left( x\right) \psi ( x) \,.
\label{nong.t.}
\end{eqnarray}
Here 
$\Omega ( x) \in G$ with $G$ being a semisimple Lie group
which is called the {gauge group}\index{gauge group}
($G={SU}(3)$ for QCD). Equation~\rf{nong.t.} 
demonstrates that $\psi$ belongs to the {fundamental} representation of $G$.

The {unitary} gauge group is when
\be
\Omega^{-1}(x)~=~\Omega^{\dagger}(x) \,,
\ee
while an extension to other Lie groups is straightforward.
%%$\Omega^{\dagger}(x)$  should be substituted  
%%in the formulas below by $\Omega^{-1}(x)$.
Then we have
\begin{eqnarray}
\psi^\dagger ( x) & 
\stackrel{{\rm g.t.}}{\longrightarrow}
& \psi^{\prime\,\dagger} ( x)~
=~\psi^\dagger \!\left( x\right) \Omega^{\dagger}\! \left( x\right).
\label{-nong.t.}
\end{eqnarray}

In analogy with QCD, the gauge group $G={SU}(\N)$ is usually associated 
with {color}\index{color} and
the proper index of $\psi$ is called the color index.
 
The gauge transformation~\rf{nong.t.} of the {matter field} $\psi$ can be  
compensated by a transformation\index{transformation!gauge!non-Abelian} 
of the non-Abelian gauge field $\A_\mu$ 
which belongs to the {adjoint} representation of $G$:
\begin{eqnarray}
\hspace*{-5mm}
\lefteqn{\A_\mu ( x) 
\stackrel{{\rm g.t.}}{\longrightarrow}
 \A_\mu^\prime ( x)}\non &&=
\Omega \!\left( x\right) \A_\mu\! \left(x\right) 
\Omega^{\dagger}\!\left(x\right) 
+\i\,\Omega \!\left( x\right) \partial _\mu
\Omega ^{\dagger}\!\left( x\right) .
\label{Ag.t.}
\end{eqnarray}

It is convenient to introduce the Hermitian
matrix
\begin{eqnarray}
\left[ \A_\mu \!\left( x\right) \right] ^{ij} & = & g \sum_a A_\mu ^a\!\left(
x\right) \left[ t^a\right] ^{ij}, 
\label{calA}
\end{eqnarray}
where $g$ is the gauge {coupling constant}.

The matrices $\left[ t^a\right] ^{ij}$ are the generators of $G$ 
($a=1,\ldots,\N^2-1$\index{generators!$SU(\N)$} 
for $SU(\N)$) which are normalized such that
\be
\tr{t^a t^b}~=~ \delta^{ab},
\label{tracetatb}
\ee
where $\tr{}$ is the trace over the matrix indices $i$ and $j$. 

Quite often another normalization of the generators
with an extra factor of $1/2$,
$\tr{\tilde t^a \tilde t^b}~=~ \frac12 \delta^{ab}$,
is used for historical reasons, in particular
$\tilde t^a~=~{\sigma^a}/{2} $
for the $SU(2)$ group,  where $\sigma^a$ are the Pauli matrices. 
This results in the redefinition of the coupling constant,
$\tilde g^2=2 g^2$.

Equation~\rf{calA} can be inverted to give
\be
A_\mu^a ( x)~=~\frac{1}{g} \tr{\A_\mu\! \left( x\right)t^a}.
\ee

Substituting
\be\displaystyle
\Omega(x)~=~\e^{\i\alpha(x)},
\ee
we obtain for an infinitesimal $\alpha$:
\bea
\delta\,\A_\mu(x)&
\stackrel{{\rm g.t.}}{=}
& \nabla_\mu^{{\rm adj}}\, \alpha(x)\,.
\label{iAg.t.}
\eea
Here
\bea
\nabla_\mu^{{\rm adj}}\, \alpha &\equiv& \partial _\mu \alpha 
- \i\left[ \A_\mu ,\alpha\right] 
\label{adjder}
\eea 
is the covariant derivative\index{covariant derivative!non-Abelian} 
in the {adjoint} representation of $G$, while
\bea
\nabla_\mu^{{\rm fun}}\, \psi &\equiv& \partial _\mu \psi
- \i\, \A_\mu \psi
\label{funder} 
\end{eqnarray} 
is that in the {fundamental} representation.
It is evident that
\bea
\nabla_\mu^{{\rm adj}}\, B(x) &=&
[\nabla_\mu^{{\rm fun}},B(x) ]\,,
\label{fader}
\eea
where $B(x)$ is a matrix-valued function of $x$.

The {QCD action}\index{quantum chromodynamics}
\index{quantum chromodynamics!Euclidean action} 
is given in the matrix notation as
\bea
\lefteqn{
S\left[ \A,\psi ,\bar \psi \right]}\non
&=& \int \d^4 x\left[ \bar \psi \gamma
_\mu \left( \partial _\mu -\i\,\A_\mu \right) \psi + m\bar\psi \psi+
\frac 1{4g^2}\tr{\F_{\mu \nu}^2}\right], \non &&
\label{QCDaction}
\eea
where
\begin{eqnarray}
\F_{\mu \nu } & = & \partial _\mu \A_\nu -\partial _\nu \A_\mu 
-\i\left[ \A_\mu,\A_\nu \right] 
\label{defF}
\end{eqnarray}
is the (Hermitian) matrix of the {non-Abelian field strength}.
\index{field strength!non-Abelian}

This action is invariant under the local gauge
transformation since\index{symmetry!gauge}
\bea
\F_{\mu \nu }(x) &
\stackrel{{\rm g.t.}}{\longrightarrow} &
\Omega(x)\, \F_{\mu \nu } (x)\,\Omega^\dagger(x) 
\label{Fg.t.}
\eea
or
\bea
\delta \F_{\mu \nu }(x) &
\stackrel{{\rm g.t.}}{=} &
-\i\, [ \F_{\mu \nu } (x) , \alpha(x) ]
\label{iFg.t.}
\eea
for the infinitesimal gauge transformation.

For the Abelian group $G={U}(1)$, the formulas 
recover those for QED. 
 
\subsection{Non-Abelian phase factors (Wilson loops)}

To compare phases of wave functions at distinct points,
a {non-Abelian} extension of the {parallel transporter} is needed.
The proper extension of the
Abelian formula~\rf{aphasefactor}: %%is
\bea
U[\Gamma_{yx}]&=& {\boldmath P} \e^{ \i\int_{\Gamma_{yx}} \d z^\mu \A_\mu(z)},
\label{phasefactor}
\eea
includes the symbol ${\boldmath P}$ of path-ordering.

Although the matrices $\A_\mu(z)$ do not commute, the path-ordered exponential
on the RHS of \eq{phasefactor}
is defined unambiguously. This is obvious after rewriting 
the phase factor in an equivalent form
\bea
{\boldmath P} \e^{\i\int_{\Gamma_{yx}} \d z^\mu \A_\mu(z)} &=&
 {\boldmath P}\e^{\i\int_0^\tau\d t \,\dot{z}^\mu(t) \,
\A_\mu\left(z(t)\right)} .
\eea

The {path-ordered exponential} in \eq{phasefactor}
can be understood as
\bea
U[\Gamma_{yx}]&=& \prod\limits_{t=0}^\tau
\left[1+ \i\, \d t\,\dot{z}^\mu(t)\, \A_\mu(z(t))\right].  
\label{phasefactorprdt}
\eea
Using \eq{dz=dtdotz}, \eq{phasefactorprdt} can also be written as
\bea 
U[\Gamma_{yx}]&=& \prod\limits_{z \in \Gamma_{yx} }
\left[1+\i\, \d z^\mu \A_\mu(z)\right].  
\label{phasefactorprd}
\eea

If the contour $\Gamma_{yx}$ is {discretized},
then the non-Abelian phase factor\index{phase factor!non-Abelian} 
is {approximated} by
\bea 
\lefteqn{U[\Gamma_{yx}]}\non &&= \lim_{M\to\infty}\prod\limits_{i=1}^M
\left[1+\i\, (z_i-z_{i-1})^\mu
\A_\mu\left(\frac{z_{i}+z_{i-1}}{2}\right)\right], \non &&
\label{dphasefactorprd}
\eea
%This formula is to be compared with \eq{product}.
which obviously reproduces~\rf{phasefactorprd} in the limit 
$z_{i-1}\to z_i$.

The non-Abelian phase 
factor\index{phase factor!non-Abelian}~\rf{phasefactor} is an
element of the {gauge group} $G$ itself, while $\A_\mu$ belongs
to the {Lie algebra} of $G$.

Matrices are rearranged in {inverse order} under {Hermitian 
conjugation}:
\bea
\hspace*{5.8mm}
 U^\dagger[{\Gamma_{yx}}]  ~=~ U[{\Gamma_{xy}}] \,.
\label{orient}
\eea
The notation $\Gamma_{yx}$ means the {orientation} of the contour 
from $x$ to $y$, while $\Gamma_{xy}$ denotes the opposite orientation from 
$y$ to $x$.  These two
 result in {opposite orders} of multiplication for the matrices
in the path-ordered product.

The phase factors obey the {backtracking} ({\em zig-zag}) 
condition\index{backtracking condition} 
\bea
U[{\Gamma_{yx}}] \, U[{\Gamma_{xy}}]&=&1\,.
\label{backtracking}
\eea

The gauge field
$\A_\mu$ in the {discretized} phase factor~\rf{dphasefactorprd}
is chosen
at the {\em center}\/ of the $i$th interval in order to satisfy 
\eq{backtracking} at finite discretization.

Under the {gauge transformation}~\rf{Ag.t.} $U[\Gamma_{yx}]$
transforms as 
\begin{eqnarray}
U[\Gamma_{yx}]&
\stackrel{{\rm g.t.}}{\longrightarrow}
& \Omega \!\left( y\right)  U\!\left[\Gamma_{yx}\right] 
\Omega ^{\dagger}\!\left( x\right).
\label{non86}
\end{eqnarray}
This formula stems from the fact that
\bea
\lefteqn{\hspace*{-5mm}
\left[1+\i\, \d z^\mu \A_\mu(z)\right]
\stackrel{{\rm g.t.}}{\longrightarrow} 
\left[1+\i\, \d z^\mu \A_\mu^\prime(z)\right]} \non &=&
\Omega(z+\d z)\left[1+ \i\,\d z^\mu \A_\mu(z)\right]
\Omega^\dagger(z)
\eea
which can be proven by substituting \eq{Ag.t.},
so that $\Omega^\dagger(z)$ and $\Omega(z)$ {\em cancel}\/ in
the definition~\rf{phasefactorprd} at the {intermediate} point $z$.

A consequence of \eq{non86} is that $\psi(x)$,
transported by the matrix $U[\Gamma_{yx}]$ 
to the point $y$, transforms under the 
{gauge transformation}\index{transformation!gauge!non-Abelian} as $\psi(y)$:
\bea
U[{\Gamma_{yx}}] \;\psi(x) &
\stackrel{{\rm g.t.}}{\sim}& ``\psi(y)\hbox{''} .
\eea
Therefore, $U[\Gamma_{yx}]$ is, indeed, a {parallel transporter}.
\index{parallel transport!non-Abelian}
 
It follows from these formulas that 
$\bar{\psi}(y)\,U[\Gamma_{yx}]\,\psi(x)$ is {gauge invariant}: 
\bea
\bar{\psi}(y)\,U[\Gamma_{yx}]\,\psi(x)
&\stackrel{{\rm g.t.}}{\longrightarrow}&
\bar{\psi}(y)\,U[\Gamma_{yx}]\,\psi(x)\,.
\eea
Another consequence of \eq{non86} is that the trace of the phase factor for
a {closed} contour $\Gamma$ is {gauge invariant}:
\bea
\tr{}{\boldmath P}\e^{\i\oint_\Gamma \d z ^\mu \A_\mu ( z ) }
&\stackrel{{\rm g.t.}}{\longrightarrow}&
\tr{}{\boldmath P}\e^{\i\oint_\Gamma \d z ^\mu \A_\mu ( z ) }. 
\label{trU}
\eea
This  is quite similar to the {Abelian} phase factor.

The {sufficient and necessary} condition for the phase
factor to be {independent} on a local variation {of the path} is 
the {\em vanishing}\/ of $\F_{\mu\nu}$. The formulas of this type are 
well-known in {differential geometry} where
parallel transport\index{parallel transport!non-Abelian} 
around a small {closed contour} determines
the {curvature}. $\F_{\mu\nu}$ in 
{Yang--Mills theory}\index{Yang--Mills theory}
is the proper {curvature} in an internal color space
while $\A_\mu$ is the {connection}. 

\subsection*{A historical remark}

An analog of the phase factor was first introduced by H.~Weyl 
in 1919~\cite{Wey19}
in his attempt to describe {gravitational} and {electromagnetic} 
interactions of electron on equal footing. 
What he did is associated in modern
language with the {scale} rather than the {gauge} transformation, 
\ie the 
vector-potential was not multiplied by \/$\i$\/ as in \eq{aphasefactor}. 
This 
explains the term ``{gauge invariance}''\index{symmetry!gauge} 
-- gauging literally means {fixing a scale}.

The factor of \/${\i}\/$ was inserted by London in 1927~\cite{Lon27} after
creation of {quantum mechanics} and the recognition
 that the electromagnetic interaction corresponds 
to the freedom of 
choice of the {phase of a wave function} 
and not to a {scale transformation}.

%\end{document}
%\vspace*{2cm}
\part{Lattice gauge theories}

{Lattice gauge theories}
 were proposed 
by Wilson in 1974~\cite{Wil74} to explain {quark confinement} in QCD. 
Lattice gauge theory is a nonperturbative {regularization} of 
gauge theory and its 
 nontrivial {definition} beyond {perturbation theory}. It provides a 
nonperturbative {quantization} of gauge fields by a lattice.
Lattice gauge theories uses an analogy between 
quantum field theory and {statistical mechanics} and 
offers a possibility 
of applying nonperturbative methods:  the {strong-coupling expansion} or
the numerical {Monte Carlo method} to QCD, 
which provide {evidence for quark confinement}.

\section{Lattice formulation}
\subsection{The lattice}

A {lattice} approximates 
{continuous} space by a {discrete} set of points.
In {Euclidean formulation} 
the lattice is along {all} four 
coordinates, while the time is left continuous in {Hamiltonian} 
approach.

The lattice\index{lattice} 
is defined as a set of points of $d$-dimensional 
{Euclidean} space with coordinates
\begin{eqnarray}
\hspace*{5.5mm}
x_\mu ~ = ~ n_\mu a \,,
\label{sites}
\eea
where the components of the vector
\bea
\hspace*{5.5mm}
n_\mu~=~(n_1,n_2,\ldots,n_d)
\eea
are {integer} numbers.
The points~\rf{sites} are called the lattice {sites}.

The dimensional constant $a$, which equals the {distance} between the 
{neighboring sites}, is called the {lattice spacing}.
{Dimensional quantities} are measured in units of $a$, 
thereby setting $a=1$. 

%\includegraphics{latt.eps} 
%A two-dimensional lattice is depicted in Fig.~\ref{FIG.1}.
\begin{figure}[b]
%%\unitlength=1.00mm
\unitlength=.80mm
\linethickness{0.6pt}
%%\hspace*{-2cm}
\hspace*{-1.8cm}
\begin{picture}(100.00,57.00)(-10,0)
\multiput(20.00,10.00)(10.00,0.00){7}{\circle*{0.60}}
\multiput(20.00,20.00)(10.00,0.00){7}{\circle*{0.60}}
\multiput(20.00,30.00)(10.00,0.00){7}{\circle*{0.60}}
\multiput(20.00,40.00)(10.00,0.00){7}{\circle*{0.60}}
\multiput(20.00,50.00)(10.00,0.00){7}{\circle*{0.60}}
\put(60.00,35.00){\vector(0,-1){4.00}}
\put(60.00,35.00){\vector(0,1){4.00}}
\put(65.00,30.00){\vector(-1,0){4.00}}
\put(65.00,30.00){\vector(1,0){4.00}}
\put(65.00,29.00){\makebox(0,0)[ct]{$a$}}
\put(59.00,35.00){\makebox(0,0)[rc]{$a$}}
\put(39.00,29.00){\makebox(0,0)[rt]{$x$}}
\put(18.00,52.00){\makebox(0,0)[rb]{\small 1}}
\put(18.00,40.00){\makebox(0,0)[rc]{\small 7}} 
\put(18.00,30.00){\makebox(0,0)[rc]{\small 8}}
\put(18.00,20.00){\makebox(0,0)[rc]{\small 9}}
\put(30.00,52.00){\makebox(0,0)[cb]{\small 2}}
\put(40.00,52.00){\makebox(0,0)[cb]{\small 3}}
\put(50.00,52.00){\makebox(0,0)[cb]{\small 4}}
\put(60.00,52.00){\makebox(0,0)[cb]{\small 5}}
\put(70.00,52.00){\makebox(0,0)[cb]{\small 6}}
\put(82.00,52.00){\makebox(0,0)[lb]{\small 1}}
\put(82.00,40.00){\makebox(0,0)[lc]{\small 7}}
\put(82.00,30.00){\makebox(0,0)[lc]{\small 8}}
\put(82.00,20.00){\makebox(0,0)[lc]{\small 9}}
\put(82.00,8.00){\makebox(0,0)[lt]{\small 1}}
\put(70.00,8.00){\makebox(0,0)[ct]{\small 6}}
\put(60.00,8.00){\makebox(0,0)[ct]{\small 5}}
\put(50.00,8.00){\makebox(0,0)[ct]{\small 4}}
\put(40.00,8.00){\makebox(0,0)[ct]{\small 3}}
\put(30.00,8.00){\makebox(0,0)[ct]{\small 2}}
\put(18.00,8.00){\makebox(0,0)[rt]{\small 1}}
\end{picture}
\caption[]{Two-dimensional lattice with periodic boundary conditions.}
   \label{fig:lat}
   \end{figure}
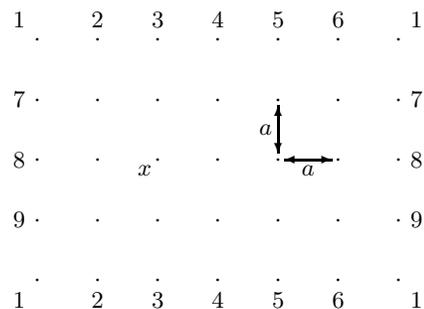
A 2d lattice with {periodic boundary conditions} is depicted in 
Fig.~\ref{fig:lat}. 
Sites\index{boundary condition!periodic}
with the same numbers are {identified}. The spatial size of the lattice 
has $L_1=6$ and $L_2=4$.      
An analogous 4d lattice is called {hypercubic}.

The next concepts are the {link\index{link}} and
{plaquette} of a lattice which are shown in Fig.~\ref{fig:lipla}.
\begin{figure}[tb]
%%\unitlength=1.00mm
\unitlength=0.80mm
\linethickness{0.6pt}
%%\hspace*{-7cm}
\hspace*{-6cm}
\begin{picture}(60.00,18.00)(-8,0)
\thicklines
\put(40.00,10.00){\circle{2.00}}
\put(41.00,10.00){\line(1,0){18.00}}
\put(60.00,10.00){\circle{2.00}}
\put(50.00,12.00){\makebox(0,0)[cb]{$l$}}
\put(41.00,5.00){\makebox(0,0)[rb]{$x$}}
\put(59.00,5.00){\makebox(0,0)[lb]{$x+a\hat\mu$}}
\end{picture}
\begin{picture}(60.00,40.00)(-8,0)
\thicklines
\put(40.00,10.00){\circle{2.00}}
\put(41.00,10.00){\line(1,0){18.00}}
\put(60.00,10.00){\circle{2.00}}
\put(60.00,11.00){\line(0,1){18.00}}
\put(40.00,30.00){\circle{2.00}}
\put(41.00,30.00){\line(1,0){18.00}}
\put(60.00,30.00){\circle{2.00}}
\put(40.00,11.00){\line(0,1){18.00}}
\put(50.00,20.00){\makebox(0,0)[cc]{$p$}}
\put(41.00,5.00){\makebox(0,0)[rb]{$x$}}
\put(59.00,5.00){\makebox(0,0)[lb]{$x+a\hat\mu$}}
\put(41.00,33.60){\makebox(0,0)[rb]{$x+a\hat\nu$}}
\put(59.00,33.60){\makebox(0,0)[lb]{$x+a\hat\mu+a\hat\nu$}}
\end{picture}
\caption[A link of a lattice]   
{Link (left) and plaquette (right).
}%}
   \label{fig:lipla}
   \end{figure}
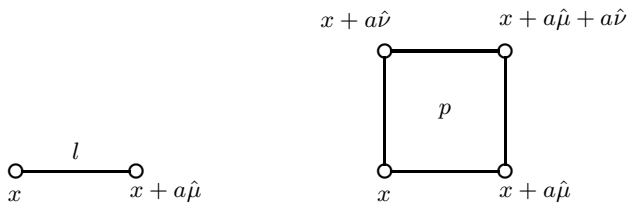
A {link} $l = \left\{ x; \mu \right\}$
 connects two neighboring sites $x$ and $x+a\hat\mu$,
where $\hat\mu$ is a unit vector along the $\mu$-direction
($\mu=1,\ldots,d$). 
A plaquette $p=\left\{ x;\mu,\nu\right\}$ is
the elementary square enclosed by four links in  the 
 directions $\mu$ and $\nu$. 
The set of four links which bound the plaquette $p$ is denoted as 
$\partial p$.
   
For an infinite lattice, the number of 
degrees of freedom = $\infty$ (but enumerable). 
To limit the this number, 
the lattice has a finite size $L_1\times L_2 \times \cdots 
\times L_d$ in all directions. 
{Periodic boundary conditions} are
usually imposed to reduce {finite-size effects}.

\subsection{Matter and gauge fields on the lattice}
%%\subsection{Lattice formulation}

{Matter field}, say a quark field, is attributed to the {\em lattice sites},
therefore a continuous field $\varphi ( x)$
is approximated by its values at the lattice sites
\begin{eqnarray}
\varphi ( x) & \Longrightarrow & \varphi_x \,.
\label{varphi_x}
\eea
The lattice field $\varphi_x$ 
is a good approximation of a continuous field 
$\varphi \left(x\right)$ when 
$a \ll$  than
the {characteristic size} of a given configuration.
A description of continuum field configuration by lattices
is illustrated by Fig.~\ref{fig:cofi}.
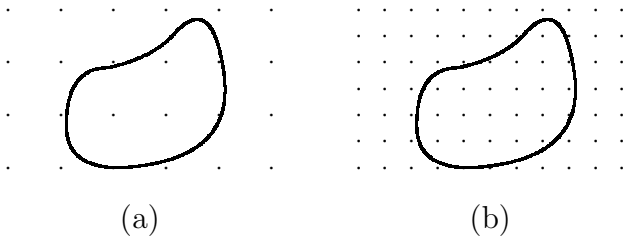
\begin{figure}[tb]
%%\unitlength=1.00mm
\unitlength=.70mm
\linethickness{0.6pt}
%\begin{picture}(50.00,33.00)(-2,0)
%%\hspace*{-2cm}
\hspace*{-1.5cm}
\begin{picture}(50.00,48.00)(-2,-15)
%\thicklines
\multiput(0.00,10.00)(10.00,0.00){6}{\circle*{0.60}}
\multiput(0.00,20.00)(10.00,0.00){6}{\circle*{0.60}}
\multiput(0.00,30.00)(10.00,0.00){6}{\circle*{0.60}}
\multiput(0.00,0.00)(10.00,0.00){6}{\circle*{0.60}}
%\put(25.00,-10.00){\makebox(0,0)[cc]{{(a)}}}
\put(25.00,-10.00){\makebox(0,0)[cc]{{\large (a)}}}
\thicklines
\bezier{80}(11.00,8.00)(11.00,19.00)(19.00,19.00)
\bezier{64}(19.00,19.00)(28.00,21.00)(32.00,26.00)
\bezier{108}(32.00,26.00)(39.00,33.00)(41.00,17.50)
\bezier{148}(21.00,0.00)(43.00,1.00)(41.00,17.50)
\bezier{80}(11.00,8.00)(11.00,0.00)(21.00,0.00)
\end{picture} 
%\begin{picture}(50.00,33.00)(-15,0)
\begin{picture}(50.00,48.00)(-17,-15)
%\thicklines
\multiput(0.00,10.00)(5.00,0.00){11}{\circle*{0.60}}
\multiput(0.00,15.00)(5.00,0.00){11}{\circle*{0.60}}
\multiput(0.00,20.00)(5.00,0.00){11}{\circle*{0.60}}
\multiput(0.00,25.00)(5.00,0.00){11}{\circle*{0.60}}
\multiput(0.00,30.00)(5.00,0.00){11}{\circle*{0.60}}
\multiput(0.00,5.00)(5.00,0.00){11}{\circle*{0.60}}
\multiput(0.00,0.00)(5.00,0.00){11}{\circle*{0.60}}
\put(25.00,-10.00){\makebox(0,0)[cc]{{\large (b)}}}
%\put(25.00,-10.00){\makebox(0,0)[cc]{{(b)}}}
\thicklines
\bezier{100}(11.00,8.00)(11.00,19.00)(19.00,19.00)
\bezier{88}(19.00,19.00)(28.00,21.00)(32.00,26.00)
\bezier{136}(32.00,26.00)(39.00,33.00)(41.00,17.50)
\bezier{170}(21.00,0.00)(43.00,1.00)(41.00,17.50)
\bezier{100}(11.00,8.00)(11.00,0.00)(21.00,0.00)
\end{picture} 
\caption[]{Description of continuum configurations by lattices.}   
\label{fig:cofi}
   \end{figure}
Lattice (a) is ``{coarse}'' and can represent the given
continuum field configuration very roughly, while lattice (b) is
``{fine''} and its spacing is small enough.

{The gauge field} is attributed to the {\em links}\/ of the lattice:
\bea
\A_\mu ( x) & \Longrightarrow &  
 U_\mu ( x) 
 %%\equiv U_\mu \left( x\right)
 \,.
\end{eqnarray}
It looks natural since a link is characterized by a coordinate and a
direction -- the same as $\A_\mu(x)$.
Sometimes the notation~$U_{x,\mu}$ is used as an alternative   
for $ U_\mu ( x)$ to emphasize that it is attributed to links.
%and $\phi_x$ is used for $\phi(x)$ a matter field on the lattice.

The link variable $U_\mu ( x)$ can be viewed as
\begin{eqnarray}
U_\mu (x ) & = & 
{\boldmath P}\e^{\i\int_x^{x+a\hat\mu }\d z ^\mu \A_\mu
\left( z \right) } ,
\label{one-link}
\end{eqnarray}
where the integral is along the link $\{x;\mu\}$.
As $a\rightarrow 0$, this yields 
\begin{eqnarray}
U_\mu (x) & \rightarrow & \e^{\i a \A_\mu ( x) } 
\label{UvsA}
\end{eqnarray}
so that $U_\mu (x)$ is expressed via the exponential of the $\mu$th
component of the vector potential at the center of the link.

Since the path-ordered integral in \eq{one-link} depends on the orientation,
links are {oriented}. The same link, 
which connects the points $x$ and $x+a\hat\mu$, can be written either
as $\{x;\mu\}$ or as $\{x+a\hat\mu;-\mu\}$. The orientation is {positive} 
for $\mu>0$ in the former case (\ie the same as the direction of the 
coordinate axis) and is {negative} in the latter case. 

The link variable  $U_\mu (x)$ is assigned 
to links with {positive} orientations.
The $U$-matrices which are assigned to links with {negative} 
orientations are given by
\bea
\hspace*{5mm}
U_{-\mu}\! \left(x+a\hat\mu \right)~=~U_{\mu }^\dagger \!\left(x\right).
\label{one-linkorient}
\eea
This is a one-link analog of \eq{orient}.

It is clear from the relation~\rf{one-link} 
between the lattice and continuum gauge variables how to construct
{lattice phase factors} (Wilson loops) -- to
construct the contours from the links of the lattice.

\subsection{The Wilson action}

An important
 role in the lattice formulation is played by the phase factor 
for the simplest closed contour on the lattice: the (oriented) boundary of a 
{plaquette}.
The {plaquette variable} is composed from 
the {link variables}~\rf{one-link} as
\begin{eqnarray}
U ({\partial p}) & = & U_{\nu}^{\dagger}\!\left( x\right) 
U_{\mu}^{\dagger} \!\left(x+a\hat\nu\right)
U_\nu\!\left(x+a\hat\mu \right) U_\mu \!\left( x\right).
\end{eqnarray}
 
The {link variable} transforms under 
the gauge transformation as
\begin{eqnarray}
U_\mu ( x ) & 
\stackrel{{\rm g.t.}}{\longrightarrow}
 & \Omega\!\left( x+a\hat\mu\right) U_\mu \!\left( x \right)\Omega
^{\dagger}\!\left( x\right) ,
\label{latgauge}
\end{eqnarray}
where the matrix $\Omega(x)$ is attributed to 
the lattice {sites}. This defines the {lattice gauge transformation}.
\index{transformation!gauge!lattice}

The plaquette variable transforms under the lattice gauge transformation as
\begin{eqnarray}
U(\partial p ) & 
\stackrel{{\rm g.t.}}{\longrightarrow}
 & \Omega (x ) \,U (\partial p )\,
\Omega^{\dagger}( x )\,.
\label{Latg.t.}
\end{eqnarray}
Its trace over the color indices is gauge invariant:
\bea
{\rm tr}\,U (\partial p ) & 
\stackrel{{\rm g.t.}}{\longrightarrow}
 & {\rm tr}\, U (\partial p ) \,.
\end{eqnarray}

This invariance  is used in 
constructing an action\index{lattice action} 
of a {lattice gauge theory}. The simplest {Wilson action} is 
\begin{eqnarray}
S_{{\rm lat}}[U] & = & 
\sum\limits_p\left[ 1-\frac 1{\N}{\rm Re\,tr\,}
U ( \partial p ) \right].
\label{Waction}
\end{eqnarray}
The summation is over all the elementary plaquettes of the lattice 
(\ie over all $x$, $\mu$, and $\nu$), regardless of their orientations.

Since a reversal of the orientation  results
 in {complex conjugation}:
\bea
{\rm tr}\, U (\partial p ) &
\stackrel{{\rm reor.}}{\longrightarrow}& 
{\rm tr}\, U^\dagger( \partial p ) ~
=~ \left[{\rm tr}\, U ({\partial p})\right]^*\!,
\eea
the action
can be rewritten as the sum over {oriented plaquettes}
\begin{eqnarray}
S_{{\rm lat}}[U] & = & 
\frac 12 \sum\limits_{{{\rm orient}}\;p}
\left[ 1-\frac 1{\N}{\rm tr\,}U({\partial p})\right].
\label{Wactionprime}
\end{eqnarray}

As $a\to0$, the lattice action~\rf{Waction} becomes 
the action 
of {continuum} gauge theory. To show this, we note
that
\begin{eqnarray}
U ({\partial p}) & \rightarrow & 
\exp{\left[\i a^2 \F_{\mu \nu}(x)+{\cal O}\!\left(a^3\right)\right] } \,,
\label{UvsF}
\end{eqnarray}
where $\F_{\mu\nu}(x)$ is {non-Abelian field strength} \rf{defF}.

In the Abelian theory, the expansion~\rf{UvsF} is easily found from the 
{Stokes theorem}. The commutator of $\A_\mu(x)$ and $\A_\nu(x)$, which 
arises in the non-Abelian case, complements the field strength to the
non-Abelian one, as is ensured by the {gauge invariance}.

A transition to the {continuum limit} is performed as 
\begin{eqnarray}
 a^4 \sum\limits_p \cdots &\stackrel {a\to0}{\longrightarrow} 
& \frac 12 \int \d^4x \sum\limits_{\mu ,\nu } \cdots \,.
\end{eqnarray}
Expanding the RHS of \eq{UvsF} in $a$, we reproduce
the continuum action
\begin{eqnarray}
S_{{\rm lat}}  & \stackrel {a\to0}{\longrightarrow}
&\frac {1}{4\N} \int \d^4x
\sum\limits_{\mu ,\nu }\,
{\rm tr}\, \F_{\mu \nu }^2(x) \,.
\label{(1.23)}
\end{eqnarray}

\subsection{The Haar measure \label{ss:H.m.}}

\index{Haar measure}
{The partition function}\index{gauge theory!lattice!partition function} 
of a pure lattice gauge theory is defined by
\begin{eqnarray}
Z\!\left( \beta \right) & = & \int \prod\limits_{x,\mu }
\d U_\mu ( x)\,\e^{-\beta S[U]} \,,
\label{Latpartition}
\end{eqnarray}
where the action is given by \eq{Waction}. 

This is the analog of a 
partition function in {statistical mechanics} at an inverse temperature 
$\beta$ given \label{p:bvsgd} by
\bea
\hspace*{5mm}
\beta ~=~\frac{\N}{g^2}\,.
\label{betavsg^2}
\eea 
This formula results from comparing \eq{(1.23)} with
the gauge-field part of the continuum action~\rf{QCDaction}.

A subtle question is what is the measure $\d U_\mu ( x )$ in 
\eq{Latpartition}. To preserve the gauge invariance at finite lattice 
spacing, the integration is over the 
{Haar measure\index{Haar measure}}\/ which is an 
{invariant group measure}.\index{group measure} 
The invariance of the Haar measure under 
multiplication by an arbitrary group element from the left or from the 
right:
\bea
\hspace*{5mm}
\d U~=~\d (\Omega U)~=~\d (U\Omega ^{\prime })\,,
\eea
guarantees the {gauge invariance} of the partition 
function~\rf{Latpartition}.

This invariance of the Haar measure is crucial for the Wilson 
formulation of lattice gauge theories.\index{Haar measure}

An explicit expression 
for the Haar measure\index{Haar measure} can be written 
for  the $SU(2)$ gauge group, 
whose element is 
parametrized by a unit four-vector $a_\mu$: 
\begin{eqnarray}
\hspace*{5mm}
U ~ = ~ a_4
{\boldmath I}+\i\vec a\vec \sigma   \qquad  a_\mu^2=1 \,,
\eea
where $\vec{\sigma}$ are the Pauli matrices. 
The Haar measure\index{Haar measure} for $SU(2)$ is
\bea 
\d U & = & \frac{1}{\pi^2} \prod\limits_{\mu =1}^4 \d a_\mu \,\delta^{(1)}
\!\left( a_\mu ^2-1\right),
\label{SU(2)Haar}
\end{eqnarray}
since $\det{U}=a_\mu^2$.

The partition function~\rf{Latpartition} characterizes vacuum effects in 
quantum theory. Physical quantities are given by the averages 
\begin{eqnarray}
\LA F[U]\,\RA 
& = & Z^{-1}( \beta ) \int \prod\limits_{x,\mu }
\d U_\mu( x )
\e^{-\beta S[U]}\,F[U]\,, 
\label{Lataverage}
\end{eqnarray}
where $F[U]$ is a gauge-invariant functional of  
$U_\mu ( x)$. These become the 
expectation values in the continuum theory as $a\to0$  and $\beta=\N/g^2$.

\section{Wilson loops on a lattice}
 
Lattice phase factors\index{phase factor!lattice} 
are associated with paths drawn on the lattice.

To write down the phase factor on the 
lattice via the link variables, let us specify the (lattice) contour $C$
by its initial point $x$ and by the directions (some may be 
negative) of the links forming the contour
\bea
\hspace*{6mm}
C~=~\{x;\mu_1,\ldots,\mu_n   \}\,.
\label{lC}
\eea
The lattice phase factor\index{phase factor!lattice}  
$U(C)$ is given by
\bea
U\!\left(C\right)&=&
U_{\mu_n}\! \left(x+a\hat{\mu}_1+\cdots+a\hat{\mu}_{n-1}\right) \cdots
\non && 
\times \,U_{\mu_2}\!\left(x+a\hat{\mu}_1\right) U_{\mu_1} \!\left( x \right).
\eea
For the links with a negative direction, it is again convenient to use
\eq{one-linkorient}.

A closed contour has $\hat{\mu}_1+\cdots+\hat{\mu}_n=0$.
The gauge 
invariant trace of the phase factor for a closed contour 
is called the {Wilson loop}.\index{Wilson loop}

The average of the Wilson loop is determined to be
\bea
W(C) &\equiv& \LA \frac 1{\N} {\rm tr}\, U(C) \RA \non
& = & Z^{-1}( \beta ) \int \prod\limits_{x,\mu }
\d U_\mu (x)
\e^{-\beta S[U]}\,\frac 1{\N}{\rm tr}\,U(C) .~~~~~
\label{Wlaverage}
\eea
This average is often called the {Wilson loop average}.
\index{Wilson loop!average of}

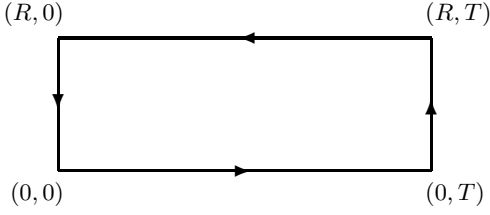
\begin{figure}[tb]
%%\unitlength=1.00mm
\unitlength=0.80mm
\linethickness{0.6pt}
\hspace*{-1.7cm}
\begin{picture}(100.00,38.00)(-10,2)
\thicklines
\put(19.00,9.00){\vector(1,0){32.00}}
\put(51.00,9.00){\line(1,0){30.00}}
\put(81.00,9.00){\vector(0,1){12.00}}
\put(81.00,21.00){\line(0,1){10.00}}
\put(81.00,31.00){\vector(-1,0){32.00}}
\put(49.00,31.00){\line(-1,0){30.00}}
\put(19.00,31.00){\vector(0,-1){12.00}}
\put(19.00,19.00){\line(0,-1){10.00}}
\put(20.00,7.00){\makebox(0,0)[rt]{$(0,0)$}}
\put(80.00,7.00){\makebox(0,0)[lt]{$(0,\T)$}}
\put(20.00,33.00){\makebox(0,0)[rb]{$(R,0)$}}
\put(80.00,33.00){\makebox(0,0)[lb]{$(R,\T)$}}
\end{picture}
\caption[Rectangular loop of size $R\times \T$]   
{Rectangular Wislon loop.
}
   \label{fig:recta}
   \end{figure}

A very important role is played by the Wilson loop averages 
for rectangular contours.
Such a contour lying in the $(x,t)$-plane is depicted in Fig.~\ref{fig:recta}.
This rectangular Wilson loop is of the size $R\times \T$.

The Wilson loop average is related for $\T\gg R$ to the 
{interaction energy} of the static (\ie infinitely heavy)
quarks, separated by a distance
$R$, by the formula
\begin{eqnarray}
W\!\left(R\times \T\right) & 
\stackrel{\T\gg R}{\propto} & \e^{-E_0( R) \T} .
\label{RT}
\end{eqnarray}

It can be proved in the axial gauge $\A_4=0$, where $U_{4}(x)=1$ 
so that only vertical
segments   contribute to
$\,U\!\left({R\times \T}\right)$. Denoting
\bea
\Psi_{ij}(t)&\equiv &
\left[{\boldmath P}\e^{\i\int_0^R \d z_1\,A_1(z_1,\ldots,t) } \right]_{ij},
\eea
we have
\bea
W\!\left(R\times \T\right)&=&\LA \frac1{\N} {\rm tr}\, \Psi\!\left(0\right)
\Psi^\dagger\!\left(\T\right)\RA.
\label{6.45}
\eea

Inserting %in \eq{6.45} 
a sum over a complete set of intermediate states 
\bea
\hspace*{5mm}
\sum\limits_n \left | n \left\rangle \right\langle n \right|~=~1\,,
\eea
we obtain
\begin{eqnarray}
W\!\left(R\times \T\right) & = & \sum\limits_n \frac 1{\N}  
\bigr\langle 
\Psi_{ij}\!\left(0\right) \bigm| n\bigr\rangle
\bigl\langle n\bigm| \Psi^\dagger_{ji}\!\left(\T\right)
\bigr\rangle   \non  & = & \sum\limits_n\frac 1{\N}
\bigl| \bigl\langle 
\Psi_{ij}\!\left(0\right) \bigm| n\bigr\rangle \bigr|^2
%{\rm tr\,}\left| \left. \left\langle {\bfP}
%\e^{\int_0^R dz_1\,A_1(z_1,\ldots,0)}
%\right| n\right\rangle \right| ^2
\e^{-E_n\T} ,
\eea
where $E_n$ is the energy of the state $\left| n \RA$.
As $\T\to\infty$, only the ground state 
with the lowest energy survives in the sum over states and
finally we find
\bea
W\!\left(R\times \T\right) &
\stackrel{{{\rm large }}~\T}
\longrightarrow  & \e^{-E_0(R)\T} .
\end{eqnarray}

Since nothing in the derivation relies on the lattice,
it holds for a rectangular loop in the continuum theory as 
well.

\section{Strong-coupling expansion \label{ss:s.c.e.}}
%%\section{Integrals over unitary group}

\index{strong-coupling expansion}
Lattice path 
integrals can be calculated either by  
{perturbation theory} in $g^2$ or  
by an expansion in 
$ \beta\propto1/g^2 $. 
This is called 
the {strong-coupling} expansion \index{strong-coupling expansion}
in analogy of the high-temperature expansion in statistical
mechanics  since $\beta$ is the analog of inverse temperature.

To perform the strong-coupling 
expansion,\index{strong-coupling expansion} we expand the 
exponential of the lattice action in $\beta$.
Then the problem is to calculate the integrals over {unitary group}:
\be
I\,{}^{i_1\cdots i_m,k_1\cdots k_n}_{j_1\cdots j_m,l_1\cdots l_n}=
\int \d U\,U_{j_1}^{i_1}\cdots U_{j_m}^{i_m}
U^\dagger{}_{l_1}^{k_1}\cdots U^\dagger{}_{l_n}^{k_n}\,,~
\label{Imn}
\ee
where the Haar measure\index{Haar measure} is normalized
\bea
\hspace*{5mm}
\int \d U~=~1\,.
\label{Haarnorm}
\eea

The integral~\rf{Imn} is 
nonvanishing only if $n=m~(\hbox{mod}~\N)$, \ie if $n=m+k \N$
with integer $k$. 

For the simplest case of $m=n=1$, the answer can easily be found by using
the unitarity of $U$ and the orthogonality relation
\begin{eqnarray}
\int \d U\,U_j^i\left. U^{\dagger}\right. _l^k 
& = & \frac 1{\N}\delta _l^i\delta _j^k \,.
\label{I11}
\end{eqnarray}

The simplest Wilson loop average is that for the loop which coincides with the 
boundary of a plaquette. It is called the {plaquette average} 
\begin{eqnarray}
W(\partial p) & = & \left\langle \frac 1{\N} 
{\rm tr\,}U({\partial p})\right\rangle .
\label{plaquetteaverage}
\eea

To calculate the plaquette average to order $\beta$, it is 
sufficient to retain only the terms ${\cal O}(\beta)$ in the expansion 
of the exponentials: 
\begin{widetext}
\be
W(\partial p) =  \frac{\displaystyle\int \prod\limits_{x,\mu }
\d U_\mu ( x)
\Big[ 1+\beta \sum\limits_{p^{\prime }}
\frac 1{\N}{\rm Re\,tr\,}U({\partial p^{\prime }})\Big] 
\frac 1{\N}{\rm tr\,}U({\partial p})}{\displaystyle\int \prod\limits_{x,\mu }
\d U_\mu ( x)
\Big[ 1+\beta
\sum\limits_{p^{\prime }}\frac 1{\N} {\rm Re\, tr\,}
U({\partial p^{\prime }})
\Big] } +{\cal O}\!\left( \beta^2\right). %%\nonumber \\*[-4mm]
\label{Wp}
\ee
\end{widetext}
The group integration can then be performed by remembering that 
\begin{eqnarray}
\int \d U_\mu ( x )\,[ U_\mu ( x )] _j^i\,
[ U_\nu^{\dagger}( y )]
_l^k & = & \frac 1{\N}\delta _{xy}\,
\delta _{\mu \nu }\,\delta _l^i\,\delta _j^k 
\label{dlinks}
\end{eqnarray}
at different links.
 
Using this property of the group integral in \eq{Wp}, we immediately see 
that the denominator is equal to $1$ (each link is 
encountered no more than once), while
the only nonvanishing contribution in the numerator is from
the plaquette $p'$, which coincides with $p$ but has the opposite orientation
as is depicted in Fig.~\ref{fig:pla2}.
\begin{figure}[tb]
%%\unitlength=1.00mm
\unitlength=0.80mm
\linethickness{0.6pt}
%%\vspace*{-.2cm}\mbox{}
\hspace*{-5cm}
\begin{picture}(60.00,24.00)(-10,10)
%\begin{picture}(60.00,30.00)(-3,10)
\thicklines
\put(51.00,10.00){\line(1,0){9.00}}
\put(40.00,10.00){\vector(1,0){11.00}}
\put(60.00,21.00){\line(0,1){9.00}}
\put(60.00,10.00){\vector(0,1){11.00}} 
\put(49.00,30.00){\line(-1,0){9.00}}
\put(60.00,30.00){\vector(-1,0){11.00}}
\put(40.00,19.00){\line(0,-1){9.00}}
\put(40.00,30.00){\vector(0,-1){11.00}} 
%p'
\put(49.00,12.00){\line(-1,0){7.00}}
\put(58.00,12.00){\vector(-1,0){9.00}}
\put(58.00,19.00){\line(0,-1){7.00}}
\put(58.00,28.00){\vector(0,-1){9.00}} 
\put(51.00,28.00){\line(1,0){7.00}}
\put(42.00,28.00){\vector(1,0){9.00}}
\put(42.00,21.00){\line(0,71){7.00}}
\put(42.00,12.00){\vector(0,1){9.00}} 
\put(61.00,20.00){\makebox(0,0)[lc]{$\partial p$}}
\put(50.00,27.00){\makebox(0,0)[ct]{$\partial p'$}}
\end{picture}
\caption[]{Boundaries of plaquettes with opposite orientations.}
   \label{fig:pla2}
   \end{figure}
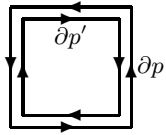
The boundaries of the plaquettes $p$ and $p'$ have opposite
orientations $\partial p$ and $\partial p'$, respectively.

The final answer for the plaquette average is
\be
\left.
\begin{array}{rcl}
\medskip
W(\partial p) &=& \displaystyle
\frac{\beta}{2\N^2}\qquad \hbox{for $SU(\N$) with }\N\geq 3\,, \\*
W(\partial p) &=& \displaystyle
\frac{\beta}{4}\qquad \hbox{ \ \ \ for $SU(2$)}\,.
\end{array}
\right\}
\label{(1.39)}
\ee
The result for $SU(2)$ differs by a factor of $1/2$ because 
tr$\,U(\partial p)$ is 
real for $SU(2)$ so that the orientation of the plaquettes can be 
ignored.

According to \eq{dlinks}, a nonvanishing result emerges only
when plaquettes, arising from the expansion of the exponentials of 
\eq{Wlaverage} in $\beta$, completely cover a surface enclosed by
the given loop $C$ as is shown in Fig.~\ref{fig:fill}.
\begin{figure}[tb]
\unitlength=1.00mm
\linethickness{0.6pt}
\hspace*{-2cm}
\begin{picture}(100.00,35.00)(-10,8)
\put(19.00,9.00){\vector(1,0){32.00}}
\put(51.00,9.00){\line(1,0){30.00}}
\put(81.00,9.00){\vector(0,1){12.00}}
\put(81.00,21.00){\line(0,1){10.00}}
\put(81.00,31.00){\vector(-1,0){16.00}}
\put(65.00,31.00){\line(-1,0){14.00}}
\put(51.00,31.00){\vector(0,1){6.00}}
\put(51.00,37.00){\line(0,1){4.00}}
\put(51.00,41.00){\vector(-1,0){17.00}}
\put(34.00,41.00){\line(-1,0){15.00}}
\put(19.00,41.00){\vector(0,-1){17.00}}
\put(19.00,24.00){\line(0,-1){15.00}}
%plaquettes
\multiput(29.00,11.00)(10.00,0.00){6}{\vector(-1,0){5.00}}
\multiput(21.00,11.00)(10.00,0.00){6}{\line(1,0){3.00}}
\multiput(21.00,19.00)(10.00,0.00){6}{\vector(1,0){5.00}}
\multiput(29.00,19.00)(10.00,0.00){6}{\line(-1,0){3.00}}
\multiput(29.00,21.00)(10.00,0.00){6}{\vector(-1,0){5.00}}
\multiput(21.00,21.00)(10.00,0.00){6}{\line(1,0){3.00}}
\multiput(21.00,29.00)(10.00,0.00){6}{\vector(1,0){5.00}}
\multiput(29.00,29.00)(10.00,0.00){6}{\line(-1,0){3.00}}
\multiput(29.00,31.00)(10.00,0.00){3}{\vector(-1,0){5.00}}
\multiput(21.00,31.00)(10.00,0.00){3}{\line(1,0){3.00}}
\multiput(21.00,39.00)(10.00,0.00){3}{\vector(1,0){5.00}}
\multiput(29.00,39.00)(10.00,0.00){3}{\line(-1,0){3.00}}
\multiput(21.00,11.00)(10.00,0.00){6}{\vector(0,1){5.00}}
\multiput(21.00,16.00)(10.00,0.00){6}{\line(0,1){3.00}}
\multiput(29.00,19.00)(10.00,0.00){6}{\vector(0,-1){5.00}}
\multiput(29.00,14.00)(10.00,0.00){6}{\line(0,-1){3.00}}
\multiput(21.00,21.00)(10.00,0.00){6}{\vector(0,1){5.00}}
\multiput(21.00,26.00)(10.00,0.00){6}{\line(0,1){3.00}}
\multiput(29.00,29.00)(10.00,0.00){6}{\vector(0,-1){5.00}}
\multiput(29.00,24.00)(10.00,0.00){6}{\line(0,-1){3.00}}
\multiput(21.00,31.00)(10.00,0.00){3}{\vector(0,1){5.00}}
\multiput(21.00,36.00)(10.00,0.00){3}{\line(0,1){3.00}}
\multiput(29.00,39.00)(10.00,0.00){3}{\vector(0,-1){5.00}}
\multiput(29.00,34.00)(10.00,0.00){3}{\line(0,-1){3.00}}
\end{picture}
\caption[Filling of a loop with plaquettes]%
{Filling of a loop with plaquettes.
}
   \label{fig:fill}
   \end{figure}
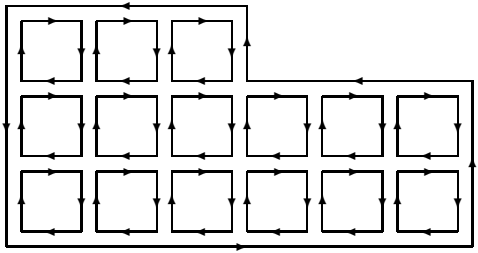
In this case each link is encountered twice (or never), once in the 
positive direction and once in the negative direction, so that all the 
group integrals are nonvanishing.

The leading order in $\beta$ corresponds to filling a {minimal 
surface}:
\bea
W(C)\!&=&\!
 \left[ W(\partial p ) \right]^{A_{{\rm min}}(C)}, 
\label{Latarealaw}
\eea
where $W(\partial p)$ is given by \eq{(1.39)} and 
$A_{{\rm min}}(C)$ is the area 
(in units of $a^2$) of the minimal surface.
For a rectangle
the minimal surface is a piece of the plane:
\begin{eqnarray}
W\!\left(R\times \T \right)\! & = & \! \left[ W(\partial p )\right] ^{R\,\T}  
\label{Wp^area}
\end{eqnarray}
to the leading order in $\beta$.

Contribution of
more complicated surfaces, do not lying  in the plane of the 
rectangle, is $W(C) \sim \beta^{{{\rm area}}}$ and is, therefore,  
suppressed  since their areas are larger than $ A_{{\rm min}}$.

\section{Area law and confinement \label{ss:a.l.}}

The exponential dependence of the Wilson loop average on the area of 
the minimal surface\index{minimal surface} (as in \eq{Latarealaw}) is called 
the {area law}.
It is customarily assumed that if an area law\index{area law} 
holds for loops of large 
area in the pure $SU(3)$ gauge theory 
then quarks are confined.\index{confinement of quarks}
In other words, there are no physical 
$\left|in\RA$ or $\LA out \right|$ quark states. This is the essence of
Wilson's {confinement criterion}.\index{confinement criterion}
The argument is that physical 
amplitudes ({\it e.g.}, the polarization operator) do not have quark
singularities when the Wilson criterion is satisfied. 

\subsection{Linear potential}

A justification for the Wilson criterion 
is based on the relationship~\rf{RT} between the Wilson loop average and 
the potential energy of interaction between static quarks.
When the area law 
\bea
W(C)& \stackrel{{\rm large}~C}
\longrightarrow& \e^{-K A_{{\rm min}}(C)}
\label{arealaw}
\eea
holds for large loops, the potential energy is a linear function of the 
distance between the quarks:
\bea
\hspace*{5mm}
E( R) ~ = ~ KR \,.
\label{KR}
\eea

\begin{figure}[tb]
%%\unitlength=1.00mm
\unitlength=.70mm
\linethickness{0.6pt}
\vspace*{-.2cm}\mbox{}
\begin{picture}(117.00,52.00)(7,0)
\thicklines
\put(20.00,30.00){\line(1,0){30.00}}
\bezier{172}(20.00,30.00)(14.00,34.00)(35.00,34.00)
\bezier{118}(20.00,30.00)(19.00,32.00)(35.00,32.00)
\bezier{172}(20.00,30.00)(14.00,26.00)(35.00,26.00)
\bezier{118}(20.00,30.00)(19.00,28.00)(35.00,28.00)
\bezier{172}(50.00,30.00)(56.00,34.00)(35.00,34.00)
\bezier{118}(50.00,30.00)(51.00,32.00)(35.00,32.00)
\bezier{172}(50.00,30.00)(56.00,26.00)(35.00,26.00)
\bezier{118}(50.00,30.00)(51.00,28.00)(35.00,28.00)
\put(90.00,30.00){\line(1,0){20.00}}
\bezier{120}(90.00,30.00)(91.00,21.00)(100.00,21.00)
\bezier{116}(90.00,30.00)(80.00,25.00)(87.00,17.00)
\bezier{104}(87.00,17.00)(92.00,12.00)(100.00,12.00)
\bezier{120}(90.00,30.00)(91.00,39.00)(100.00,39.00)
\bezier{116}(90.00,30.00)(80.00,35.00)(87.00,43.00)
\bezier{104}(87.00,43.00)(92.00,48.00)(100.00,48.00)
\bezier{120}(110.00,30.00)(109.00,21.00)(100.00,21.00)
\bezier{116}(110.00,30.00)(120.00,25.00)(113.00,17.00)
\bezier{104}(113.00,17.00)(108.00,12.00)(100.00,12.00)
\bezier{120}(110.00,30.00)(109.00,39.00)(100.00,39.00)
\bezier{116}(110.00,30.00)(120.00,35.00)(113.00,43.00)
\bezier{104}(113.00,43.00)(108.00,48.00)(100.00,48.00)
%\put(100.00,3.00){\makebox(0,0)[cc]{{(b)}}}
%\put(35.00,3.00){\makebox(0,0)[cc]{{(a)}}}
\put(100.00,3.00){\makebox(0,0)[cc]{{\large (b)}}}
\put(35.00,3.00){\makebox(0,0)[cc]{{\large (a)}}}
\put(20.00,30.00){\circle*{1.00}}
\put(50.00,30.00){\circle*{1.00}}
\put(110.00,30.00){\circle*{1.00}}
\put(90.00,30.00){\circle*{1.00}}
\end{picture}
\caption[Lines of force between static quarks]%   
{Lines of force between static quarks.
}%}
   \label{fig:poten}
   \end{figure}
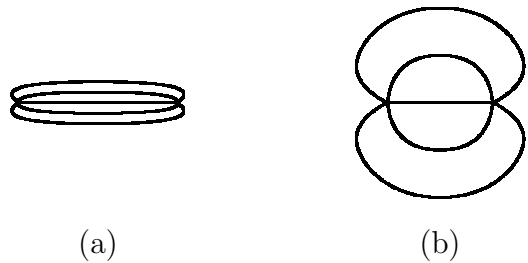
Lines of force between static quarks are depicted in Fig.~\ref{fig:poten}
for (a) linear and\index{Coulomb law}
(b) Coulomb interaction potentials. For the linear potential the lines of 
force are contracted into a tube, while they are distributed over the
whole space for the Coulomb one.

The coefficient $K$ is called the {string 
tension because the gluon field between quarks
contracts to a tube or string, whose energy is proportional to its 
length.
The value of $K$ is the energy of the string per unit length.
This string is stretched with the distance between quarks and 
prevents them from moving apart to macroscopic distances.

Equation~\rf{Wp^area} gives
\bea 
\hspace*{6mm}
K ~ = ~ \frac 1{a^2}\ln \frac{2\N^2}\beta 
~=~ \frac 1{a^2}\ln{\left(2\N g^2\right)} 
\label{Kstrongcoupling}
\end{eqnarray}
for the string tension\index{string tension} to the leading order of 
the strong-coupling expansion.
\index{strong-coupling expansion!lattice gauge theory!string tension}
The next orders of the strong-coupling expansion result in corrections in 
$\beta$ to this formula.
Confinement\index{confinement of quarks} 
holds in the lattice gauge theory to any order of the 
strong-coupling expansion.
\index{strong-coupling expansion!lattice gauge theory}

\subsection{Asymptotic scaling \label{ss:AS}}

\index{asymptotic scaling}
Equation~\rf{Kstrongcoupling} establishes the relationship between 
 lattice spacing $a$ and the coupling $g^2$. Let  $K$ equals its 
experimental value 
\bea
\hspace*{5mm}
 K~=~(400~ \hbox{MeV})^2~\approx~ 1~\hbox{GeV/fm} 
\label{400MeV}
\eea
which results from 
the slope of the Regge trajectory $\alpha^\prime=1/2\pi K$.
The slope $\alpha^\prime=1$~GeV${}^{-2}$  from 
the $\rho$~--~$A_2$~--~$g$ trajectory.

The renormalizability prescribes that variations of $a$, which
plays the role of a lattice cutoff, and of the bare charge $g^2$
should be made simultaneously in order that $K$ does not change.

Given \eq{Kstrongcoupling}, this procedure calls for $a\to\infty$ as 
$g^2\to\infty$. In other words, the lattice spacing is large in the 
strong-coupling limit, compared with $1$~fm.  
 Such a coarse lattice cannot describe 
the continuum limit and, in particular, the rotational symmetry.

In order to pass to the continuum, 
the lattice spacing $a$ should be decreased. Equation~\rf{Kstrongcoupling}
shows that $a$ decreases with decreasing $g^2$. However, this formula
ceases to be applicable in the intermediate region of $g^2\sim 1$ and,
therefore, $a\sim1$~fm.

To further decrease $a$, we further decrease $g^2$.  
While no analytic formulas are 
available at intermediate values of $g^2$, the expected relation 
between $a$ and $g^2$ for small $g^2$ is predicted by the known two-loop 
Gell-Mann--Low function of QCD.

For pure $SU(3)$ Yang-Mills, \eq{Kstrongcoupling} is replaced at small $g^2$ 
by
\begin{eqnarray}
K & = & \hbox{const}\,\cdot \frac{1}{a^2} 
\left( \frac{8\pi ^2}{11g^2}\right) 
^{\frac{102}{121}}\e^{-{8\pi ^2}/{11g^2}},
\label{AS}
\end{eqnarray}
where the two-loop Gell-Mann--Low function is used. 

The exponential dependence of $K$ on $1/g^2$ is called  
{asymptotic scaling}.\index{asymptotic scaling} 
Asymptotic scaling sets in for some 
value of $1/g^2$.
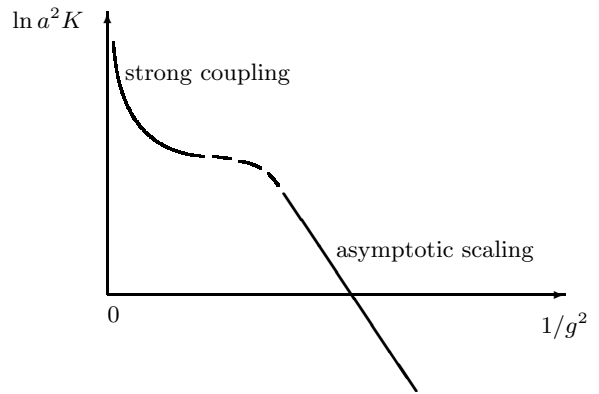
\begin{figure}[tb]
%\unitlength=1mm
\unitlength=0.8mm
\linethickness{0.6pt}
\hspace*{-3cm}
\begin{picture}(116.00,67.00)(-5,5)
\put(40.00,23.00){\vector(1,0){76.00}}
\put(40.00,23.00){\vector(0,1){47.00}}
\put(116.00,20.00){\makebox(0,0)[ct]{$1/g^2$}}
\thicklines
\bezier{160}(41.00,65.00)(42.00,47.00)(56.00,46.00)
%\bezier{34}(57.60,46.20)(58.40,46.30)(60.40,47.00)   %%with peak
%\bezier{34}(62.00,47.50)(64.00,48.50)(65.50,47.50)
%\bezier{34}(66.80,46.50)(68.30,45.00)(68.4,44.60)
%\put(69.50,43.40){\line(2,-3){24.00}}
%\put(69.40,43.40){\line(2,-3){24.00}}
\bezier{34}(57.60,46.00)(58.40,45.90)(60.40,45.60)
\bezier{34}(62.00,45.20)(64.00,44.90)(65.50,44.00)
\bezier{34}(66.80,43.10)(68.30,41.40)(68.4,41.10)
\put(69.40,39.90){\line(2,-3){22.00}}
\put(69.30,39.90){\line(2,-3){22.00}}
\put(36.00,69.00){\makebox(0,0)[rc]{$\ln a^2K$}}
%\put(36.00,47.00){\makebox(0,0)[rc]{$\ln a^2K$}} %% center
\put(78.00,28.50){\makebox(0,0)[lb]{asymptotic scaling}}
\put(43.00,58.00){\makebox(0,0)[lb]{strong coupling}}
\put(40.00,21.00){\makebox(0,0)[lt]{$0$}}
\end{picture}
\caption[]{String tension versus $1/g^2$}.%
\label{fig:ascal}
   \end{figure}
{The dependence of the string tension\index{string tension} 
on $1/g^2$ is shown in Fig.~\ref{fig:ascal}. The strong-coupling 
formula \rf{Kstrongcoupling} holds for small $1/g^2$. 
The asymptotic-scaling\index{asymptotic scaling} 
formula~\rf{AS} sets in for large $1/g^2$. Both formulas are 
not applicable in the intermediate region $1/g^2\sim1$. 
}

For such values of $g^2$, 
where asymptotic scaling\index{asymptotic scaling} holds,
the lattice gauge theory has a continuum limit.

The knowledge of the two asymptotic behaviors 
says nothing about the behavior
of $a^2K$ in the intermediate region of $g^2\sim 1$.
There can be either a smooth transition between these two regimes 
or a phase transition.
Numerical methods were introduced to study this problem.

\subsection{Relation to second-order phase transition}
 
Continuum limits of a 
lattice system are reached at the points of 
second-order phase\index{phase transition!second-order} 
transitions when the correlation length\index{correlation length} 
becomes {\em infinite}\/ in lattice 
units. 

The correlation length\index{correlation length} 
is inversely proportional to $\Lambda_{{\rm QCD}}$:
\be
\xi~\sim~\Lambda_{{\rm QCD}}^{-1}~=~{a} \exp{\left[\int 
\frac{\d g^2}{\Beta(g^2)}\right]}\,,
\label{corlenght}
\ee
where $\Beta(g^2)$ is the Gell-Mann--Low function.
The only chance for the RHS  
to diverge is to have a zero of $\Beta(g^2)$\index{Gell-Mann--Low function} 
at some fixed point $g^2=g^2_*$.\index{fixed point}
Therefore, the bare coupling should approach the fixed-point value $g^2_*$ 
to describe the continuum.

$\Beta(0)=0$ for a non-Abelian gauge theory
so that $g_*^2=0$ is a fixed-point value of the coupling constant.
Therefore, the continuum limit is associated with $g^2\to0$.

%%\end{document}

%\vspace*{4cm}
\part{$1/N$ Expansion}

An effective coupling constant of QCD becomes large at
large distances, so 
fluctuations of 
scales of different orders of magnitude 
are essential and there is no small parameter.  
't~Hooft proposed in 1974~\cite{Hoo74}
to use  the number of colors
\index{color} $\N$ of the gauge
group $SU(\N)$ as such a parameter and to perform an expansion in $1/\N$.
The motivation was the $1/N$-expansion in statistical mechanics.

The $1/\N$-expansion of QCD rearranges perturbation theory in a way consistent 
with a string\index{string representation!QCD} 
picture. The accuracy of
large-$N$\index{multicolor QCD} 
QCD is  of the order of the ratios of meson widths to their
masses (10--15\%). 
While QCD is simplified in the large-$\N$ limit, it is  
not yet solved.

\section{Index or ribbon graphs\label{ss:ribbon}} 

In order to describe the 
$1/\N$-expansion of QCD,
it is convenient to use the matrix-field representation
\begin{eqnarray}
\left[ A_\mu ( x) \right] ^{ij} & = & \sum_a A_\mu ^a(
x) \left[ t^a\right] ^{ij} .
\label{matA}
\end{eqnarray}
The matrix~\rf{matA} is Hermitian and differs from~\rf{calA} by $g$.

The propagator of the matrix field $A^{ij}(x)$ reads
%%takes the form 
\begin{equation}
\left\langle A_\mu ^{ij}( x) \, A_\nu ^{kl}( y)
\right\rangle_{\rm Gauss}  =  
\left(\delta ^{il}\delta ^{kj} -\frac 1{\N}\delta ^{ij}\delta ^{kl}\right)
D_{\mu\nu}\! \left( x-y \right),
\label{matpropagator}
\end{equation}
where we have assumed, as usual, a gauge-fixing 
%%is added to the action~\rf{QCDaction} 
to define the gluon propagator\index{propagator!gluon} 
in perturbation theory. For instance, one has%
\footnote{Here and in \eq{spacepart} $\delta_{\mu\nu}$ is 
due to Euclidean metric and is to be substituted
by $- g_{\mu\nu}$ in Minkowski space.}
\bea
D_{\mu\nu}\! \left( x-y \right)&=&\frac{1}{4 \pi^2}
\frac{\delta_{\mu\nu}}{\left( x-y \right)^2}
\label{D(x-y)}
\eea
in the Feynman gauge.

Equation~\rf{matpropagator} can be derived immediately from 
the standard formula
\be
\left\langle  A_\mu ^a( x) \, A_\nu ^b( y)
\right\rangle_{\rm Gauss}  =  \delta ^{ab} D_{\mu\nu}\! \left( x-y \right)
\ee
multiplying by the generators of the $SU( \N)$ gauge group according
to the definition~\rf{matA} and using 
the {completeness condition}\index{completeness condition!$SU(\N)$}
\be
%%\hspace*{-3mm}
\sum\limits_{a=1}^{\N^2-1}
\left( t^a\right) ^{ij}\left( t^a\right) ^{kl}  =  
\left(\delta ^{il}\delta
^{kj}-\frac 1{\N}\delta ^{ij}\delta ^{kl}\right) ~~~
\fbox{for \ $SU(\N)$}~.~~
\label{completeness}
\ee
%%where the factor of $1/2$ is due to the normalization~\rf{tracetatb}.
Alternatively, 
\eq{matpropagator} can be derived  
directly from a path integral over the matrix fields.

We concentrate  only on the structure of diagrams
in the index space, \ie the space of the indices associated with the 
$SU(\N)$ group. We shall not consider, in most cases, 
space-time structures of diagrams which are prescribed by Feynman's rules. 

Omitting at large $\N$ the second
term in parentheses on the RHS of \eq{matpropagator},  we depict the
propagator by the {\em double line}\index{double-line representation} 
\be
\left\langle A_\mu ^{ij}( x) \,A_\nu ^{kl}( y)
\right\rangle_{\rm Gauss} ~ \propto ~ 
 \delta ^{il}\delta ^{kj} ~=~ \hspace*{-6mm}
\unitlength=1mm
\linethickness{0.6pt}
\begin{picture}(31.00,4.00)(10,9)
%\begin{picture}(21.00,12.00)(0,6)
\put(20.00,11.00){\vector(1,0){6.00}}
\put(26.00,11.00){\line(1,0){4.00}}
\put(30.00,9.00){\vector(-1,0){6.00}}
\put(24.00,9.00){\line(-1,0){4.00}}
\put(18.00,12.00){\makebox(0,0)[cc]{$i$}}
\put(18.00,8.00){\makebox(0,0)[cc]{$j$}}
\put(32.00,12.00){\makebox(0,0)[cc]{$l$}}
\put(32.00,8.00){\makebox(0,0)[cc]{$k$}}
\end{picture} \hspace*{-5mm}.
\label{doubleline}
\ee
Each line, often termed the {index line},\index{index line} 
represents the {Kronecker} delta-symbol and has an orientation
which is indicated by arrows. This notation is obviously consistent
with the space-time structure of the propagator that describes
a propagation from $x$ to $y$.

{Arrows} are a result of the fact that the matrix $A_\mu^{ij}$ is 
{Hermitian} and its off-diagonal components are {complex conjugate}.

Double lines
\index{double-line representation}
appear generically in all models describing 
{matrix}\index{matrix field} fields
in contrast to {vector} (in internal symmetry space) fields, whose
propagators are depicted by single lines.  

{The three-gluon vertex} is depicted in 
the double-line\index{double-line representation} notation as 
\bea
&&\hspace*{-10mm}\mbox{\beginpicture
\setcoordinatesystem units <0.50000cm,0.50000cm>
\unitlength=0.50000cm
\linethickness=1pt
\setplotsymbol ({\makebox(0,0)[l]{\tencirc\symbol{'160}}})
\setshadesymbol ({\thinlinefont .})
\setlinear
%
% Fig POLYLINE object
%
\linethickness= 0.500pt
\setplotsymbol ({\thinlinefont .})
\plot  5.080 21.273  6.191 20.637 /
%
% Fig POLYLINE object
%
\linethickness= 0.500pt
\setplotsymbol ({\thinlinefont .})
\putrule from  4.763 21.749 to  4.763 22.860
%
% Fig POLYLINE object
%
\linethickness= 0.500pt
\setplotsymbol ({\thinlinefont .})
\plot  4.763 21.749  3.651 21.114 /
%
% Fig POLYLINE object
%
\linethickness= 0.500pt
\setplotsymbol ({\thinlinefont .})
\putrule from  5.397 21.749 to  5.397 22.860
%
% Fig POLYLINE object
%
\linethickness= 0.500pt
\setplotsymbol ({\thinlinefont .})
\plot  5.397 21.749  6.509 21.114 /
%
% Fig POLYLINE object
%
\linethickness= 0.500pt
\setplotsymbol ({\thinlinefont .})
\plot  5.857 21.495  6.079 21.368 /
%
% arrow head
%
\plot  5.827 21.439  6.079 21.368  5.890 21.549 /
%
%
% Fig POLYLINE object
%
\linethickness= 0.500pt
\setplotsymbol ({\thinlinefont .})
\plot  4.174 21.416  4.396 21.543 /
%
% arrow head
%
\plot  4.207 21.362  4.396 21.543  4.144 21.473 /
%
%
% Fig POLYLINE object
%
\linethickness= 0.500pt
\setplotsymbol ({\thinlinefont .})
\plot  4.619 21.018  4.396 20.892 /
%
% arrow head
%
\plot  4.585 21.073  4.396 20.892  4.648 20.962 /
%
%
% Fig POLYLINE object
%
\linethickness= 0.500pt
\setplotsymbol ({\thinlinefont .})
\plot  5.730 20.908  5.508 21.035 /
%
% arrow head
%
\plot  5.760 20.965  5.508 21.035  5.697 20.854 /
%
%
% Fig POLYLINE object
%
\linethickness= 0.500pt
\setplotsymbol ({\thinlinefont .})
\putrule from  5.397 22.274 to  5.397 22.115
%
% arrow head
%
\plot  5.334 22.369  5.397 22.115  5.461 22.369 /
%
%
% Fig POLYLINE object
%
\linethickness= 0.500pt
\setplotsymbol ({\thinlinefont .})
\putrule from  4.763 22.384 to  4.763 22.543
%
% arrow head
%
\plot  4.826 22.289  4.763 22.543  4.699 22.289 /
%
%
% Fig POLYLINE object
%
\linethickness= 0.500pt
\setplotsymbol ({\thinlinefont .})
\putrule from  9.842 21.749 to  9.842 22.860
%
% Fig POLYLINE object
%
\linethickness= 0.500pt
\setplotsymbol ({\thinlinefont .})
\plot  9.842 21.749  8.731 21.114 /
%
% Fig POLYLINE object
%
\linethickness= 0.500pt
\setplotsymbol ({\thinlinefont .})
\plot 10.478 21.749 11.589 21.114 /
%
% Fig POLYLINE object
%
\linethickness= 0.500pt
\setplotsymbol ({\thinlinefont .})
\plot 10.160 21.273  9.049 20.637 /
%
% Fig POLYLINE object
%
\linethickness= 0.500pt
\setplotsymbol ({\thinlinefont .})
\putrule from  9.842 22.305 to  9.842 22.147
%
% arrow head
%
\plot  9.779 22.401  9.842 22.147  9.906 22.401 /
%
%
% Fig POLYLINE object
%
\linethickness= 0.500pt
\setplotsymbol ({\thinlinefont .})
\putrule from 10.478 22.337 to 10.478 22.496
%
% arrow head
%
\plot 10.541 22.242 10.478 22.496 10.414 22.242 /
%
%
% Fig POLYLINE object
%
\linethickness= 0.500pt
\setplotsymbol ({\thinlinefont .})
\plot  9.445 21.526  9.222 21.400 /
%
% arrow head
%
\plot  9.411 21.581  9.222 21.400  9.474 21.470 /
%
%
% Fig POLYLINE object
%
\linethickness= 0.500pt
\setplotsymbol ({\thinlinefont .})
\plot  5.080 21.273  3.969 20.637 /
%
% Fig POLYLINE object
%
\linethickness= 0.500pt
\setplotsymbol ({\thinlinefont .})
\putrule from 10.478 21.749 to 10.478 22.860
%
%
% Fig POLYLINE object
%
\linethickness= 0.500pt
\setplotsymbol ({\thinlinefont .})
\plot 11.081 21.416 10.859 21.543 /
%
% arrow head
%
\plot 11.111 21.473 10.859 21.543 11.048 21.362 /
%
%
% Fig POLYLINE object
%
\linethickness= 0.500pt
\setplotsymbol ({\thinlinefont .})
\plot  9.620 20.972  9.842 21.099 /
%
% arrow head
%
\plot  9.653 20.918  9.842 21.099  9.590 21.028 /
%
%
% Fig POLYLINE object
%
\linethickness= 0.500pt
\setplotsymbol ({\thinlinefont .})
\plot 10.160 21.273 11.271 20.637 /
%
% Fig POLYLINE object
%
\linethickness= 0.500pt
\setplotsymbol ({\thinlinefont .})
\plot 10.651 20.987 10.873 20.860 /
%
% arrow head
%
\plot 10.621 20.931 10.873 20.860 10.684 21.041 /
%
%
% Fig TEXT object
%
\put{$\scriptstyle i_1$} [B] at  5.765 22.860
%\put{$\scriptstyle i_1$} [B] at  5.715 22.860
%
% Fig TEXT object
%
\put{$\scriptstyle j_1$} [B] at  4.445 22.860
%
% Fig TEXT object
%
\put{$\scriptstyle i_3$} [B] at  6.032 20.161
%
% Fig TEXT object
%
\put{$\scriptstyle j_2$} [B] at  4.128 20.161
%
% Fig TEXT object
%
\put{$\scriptstyle i_2$} [B] at  3.334 21.273
%
% Fig TEXT object
%
\put{$\scriptstyle i_1$} [B] at  9.525 22.860
%
% Fig TEXT object
%
\put{$\scriptstyle j_1$} [B] at 10.845 22.860
%\put{$\scriptstyle j_1$} [B] at 10.795 22.860
%
% Fig TEXT object
%
\put{$\scriptstyle i_2$} [B] at  9.207 20.161
%
% Fig TEXT object
%
\put{$\scriptstyle j_2$} [B] at  8.414 21.273
%
% Fig TEXT object
%
\put{$\scriptstyle j_3$} [B] at  6.826 21.273
%
% Fig TEXT object
%
\put{$\scriptstyle j_3$} [B] at 11.113 20.161
%
% Fig TEXT object
%
\put{$\scriptstyle i_3$} [B] at 11.906 21.273
%
% Fig TEXT object
%
\put{$-$} [B] at  7.620 21.749
\linethickness=0pt
\putrectangle corners at  3.234 23.165 and 13.032 20.104
\endpicture}

\nonumber \\*[1mm] && %% \hspace{-5cm}
\propto ~g \left( 
\delta^{i_1j_3}\delta^{i_2j_1}\delta^{i_3j_2} -
\delta^{i_1j_2}\delta^{i_2j_3}\delta^{i_3j_1}
\right),
\label{threegluon}
\eea
%\bea
%&&\hspace*{-6mm}
%\input{3vert}
%\nonumber \\*[-3mm] &&
%\label{threegluon}
%\eea
where the subscripts $1$, $2$ or $3$ refer to each of the three gluons.
The relative minus sign arises from the commutator in the cubic-in-$A$ term
in the action~\rf{QCDaction}. 
The color part 
is antisymmetric under an interchange of gluons. The 
(momentum space) space-time part
\bea
%%\hspace*{3.5mm}
\lefteqn{\hspace*{-5mm}
\gamma_{\mu_1\mu_2\mu_3}\!\left(p_1,p_2,p_3 \right)
= %%}\non &=&
\delta_{\mu_1\mu_2}
\left(p_1-p_2 \right)_{\mu_3}}\non &&+\delta_{\mu_2\mu_3}
\left(p_2-p_3 \right)_{\mu_1} +\delta_{\mu_1\mu_3}
\left(p_3-p_1 \right)_{\mu_2}
\label{spacepart}
\eea
is also antisymmetric. %% as well. 
We consider all three gluons as incoming
so their momenta obey $p_1+p_2+p_3=0$.
The full vertex is symmetric as is prescribed by Bose statistics.

{The four-gluon vertex}\index{vertex!four-gluon} 
involves six terms -- each of them is depicted
by a cross --
which differ by interchanging of the color indices. 
We depict the color structure of the four-gluon vertex for 
simplicity in the case when
$i_1=j_2=i$, $i_2=j_3=j$, $i_3=j_4=k$, $i_4=j_1=l$, but $i,j,k,l$ 
take on different values. Then only the following term is left:
\be
%\hspace*{4mm}
\mbox{\beginpicture
\setcoordinatesystem units <0.50000cm,0.50000cm>
\unitlength=0.50000cm
\linethickness=1pt
\setplotsymbol ({\makebox(0,0)[l]{\tencirc\symbol{'160}}})
\setshadesymbol ({\thinlinefont .})
\setlinear
%
% Fig POLYLINE object
%
\linethickness= 0.500pt
\setplotsymbol ({\thinlinefont .})
\putrule from  5.397 21.749 to  5.397 22.860
%
% Fig POLYLINE object
%
\linethickness= 0.500pt
\setplotsymbol ({\thinlinefont .})
\putrule from  5.397 22.274 to  5.397 22.115
%
% arrow head
%
\plot  5.334 22.369  5.397 22.115  5.461 22.369 /
%
%
% Fig POLYLINE object
%
\linethickness= 0.500pt
\setplotsymbol ({\thinlinefont .})
\putrule from  4.763 22.384 to  4.763 22.543
%
% arrow head
%
\plot  4.826 22.289  4.763 22.543  4.699 22.289 /
%
%
% Fig POLYLINE object
%
\linethickness= 0.500pt
\setplotsymbol ({\thinlinefont .})
\putrule from  4.763 21.114 to  3.651 21.114
%
% Fig POLYLINE object
%
\linethickness= 0.500pt
\setplotsymbol ({\thinlinefont .})
\putrule from  4.763 21.749 to  3.651 21.749
%
% Fig POLYLINE object
%
\linethickness= 0.500pt
\setplotsymbol ({\thinlinefont .})
\putrule from  6.509 21.114 to  5.397 21.114
%
% Fig POLYLINE object
%
\linethickness= 0.500pt
\setplotsymbol ({\thinlinefont .})
\putrule from  6.509 21.749 to  5.397 21.749
%
% Fig POLYLINE object
%
\linethickness= 0.500pt
\setplotsymbol ({\thinlinefont .})
\putrule from  4.763 20.003 to  4.763 21.114
%
% Fig POLYLINE object
%
\linethickness= 0.500pt
\setplotsymbol ({\thinlinefont .})
\putrule from  5.397 20.003 to  5.397 21.114
%
% Fig POLYLINE object
%
\linethickness= 0.500pt
\setplotsymbol ({\thinlinefont .})
\putrule from  5.397 20.527 to  5.397 20.369
%
% arrow head
%
\plot  5.334 20.623  5.397 20.369  5.461 20.623 /
%
%
% Fig POLYLINE object
%
\linethickness= 0.500pt
\setplotsymbol ({\thinlinefont .})
\putrule from  4.763 20.637 to  4.763 20.796
%
% arrow head
%
\plot  4.826 20.542  4.763 20.796  4.699 20.542 /
%
%
% Fig POLYLINE object
%
\linethickness= 0.500pt
\setplotsymbol ({\thinlinefont .})
\putrule from  4.128 21.749 to  4.286 21.749
%
% arrow head
%
\plot  4.032 21.685  4.286 21.749  4.032 21.812 /
%
%
% Fig POLYLINE object
%
\linethickness= 0.500pt
\setplotsymbol ({\thinlinefont .})
\putrule from  4.763 21.749 to  4.763 22.860
%
% Fig POLYLINE object
%
\linethickness= 0.500pt
\setplotsymbol ({\thinlinefont .})
\putrule from  5.874 21.114 to  5.715 21.114
%
% arrow head
%
\plot  5.969 21.177  5.715 21.114  5.969 21.050 /
%
%
% Fig TEXT object
%
\put{$\propto ~ g^2\,,$} [lB] at  7.938 21.331
%\put{$\propto ~ g^2\,,$} [lB] at  7.938 21.431
%
% Fig POLYLINE object
%
\linethickness= 0.500pt
\setplotsymbol ({\thinlinefont .})
\putrule from  5.874 21.749 to  6.032 21.749
%
% arrow head
%
\plot  5.779 21.685  6.032 21.749  5.779 21.812 /
%
%
% Fig POLYLINE object
%
\linethickness= 0.500pt
\setplotsymbol ({\thinlinefont .})
\putrule from  4.128 21.114 to  3.969 21.114
%
% arrow head
%
\plot  4.223 21.177  3.969 21.114  4.223 21.050 /
%
%
% Fig TEXT object
%
\put{$\scriptstyle i$} [B] at  5.715 22.860
%
% Fig TEXT object
%
\put{$\scriptstyle l$} [B] at  4.445 22.860
%
% Fig TEXT object
%
\put{$\scriptstyle i$} [B] at  6.826 21.907
%
% Fig TEXT object
%
\put{$\scriptstyle k$} [B] at  4.445 19.844
%
% Fig TEXT object
%
\put{$\scriptstyle j$} [B] at  5.715 19.844
%
% Fig TEXT object
%
\put{$\scriptstyle l$} [B] at  3.334 21.907
%
% Fig TEXT object
%
\put{$\scriptstyle k$} [B] at  3.334 20.796
%
% Fig TEXT object
%
%\put{\SetFigFont{6}{7.2}{rm}$j$} [B] at  6.826 20.796
\put{$\scriptstyle j$} [B] at  6.826 20.796
\linethickness=0pt
\putrectangle corners at  3.109 23.089 and  7.938 19.768
\endpicture}
 %% \nonumber %\\*[-1.35cm] \mbox{} 
\label{fourgluon}
%\\*[.5mm] \mbox{}\nonumber 
\ee
where there are no delta-symbols  since the color structure is fixed.
We pick up only one color structure by 
equating indices pairwise. 

\begin{figure}
\vspace*{3mm}
\centering{
\input{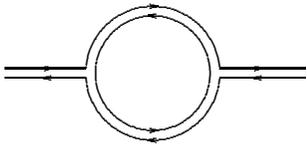}
}
\caption[]{Double-line representation of a one-loop diagram  
for gluon propagator.}
   \label{fi:1loop}
\end{figure}

Diagrams of perturbation theory can now be completely rewritten in
the {double-line} notation.\index{double-line representation}
The simplest one describing the one-loop correction to the {gluon
propagator} is depicted in Fig.~\ref{fi:1loop}.
The sum over the $\N$ indices is associated 
with the closed index line.\index{index line} 
The contribution of this diagram is $\sim g^2 \N\sim 1$.

In order for the large-$\N$ limit to be nontrivial, 
the bare coupling constant $g^2$ should satisfy 
\be
%\hspace*{5mm}
g^2~\sim~\frac 1{\N} \,.
\label{orderg}
\ee
This dependence on $\N$ is also prescribed by the {asymptotic-freedom} 
formula
\index{asymptotic freedom}
\begin{eqnarray}
g^2
 & = & \frac {12\pi^2}{11\N\ln \left( \Lambda/\Lambda _{\rm QCD}\right)}
\label{NcAF}
\end{eqnarray}
of the pure $SU(\N)$ gauge theory.

Thus, the contribution of the diagram in Fig.~\ref{fi:1loop} is of order
 \ $\sim~g^2 \N~\sim~1$ \
in the large-$\N$ limit.

The double lines \index{double-line representation} can be viewed
as bounding a piece of a plane. These lines represent
a {two-dimensional} object. In mathematics these double-line graphs 
are often called {ribbon} graphs\index{ribbon graph} 
or {fatgraphs}.\index{fatgraph, {\it see} ribbon graph} 
They are connected with 
{Riemann} surfaces.\index{Riemann surface}

\subsection*{Remark on the $U(\N)$ gauge group}

The double-line representation of\index{double-line representation} 
the diagrams holds,
strictly speaking, only for the $U(\N)$ gauge group, whose  
 generators\index{generators!$U(\N)$} 
\be
%\hspace*{5mm}
T^A=\left(t^a,\,\boldmath{I}/{\sqrt{\N}} \right) ,~~
\tr T^A T^B= \delta^{AB}  \quad \fbox{$A=1,\ldots,\N^2$}\,.
\ee
obey the completeness condition\index{completeness condition!$U(\N)$} 
\be
\hspace*{-1mm}
\sum\limits_{A=1}^{\N^2}
\left( T^A\right) ^{ij}\left( T^A\right) ^{kl}  =  
\delta ^{il}\delta^{kj} ~~~~
\fbox{for \ $U(\N)$}\,.
\label{completenessU}
\ee
Elements of both  $SU(\N)$ and $U(\N)$ can be represented in the form
\be
%\hspace*{5mm}
U~=~\e^{\i B},
\ee 
where $B$ is a general Hermitian matrix for $U(\N)$ and 
a traceless Hermitian matrix for $SU(\N)$.

The large-$\N$ limit of both the $U(\N)$ and $SU(\N)$ groups is 
the same.

\section{Planar and nonplanar graphs\label{ss:pnp}}

The double-line representation\index{double-line representation}
of perturbation-theory diagrams %%in the index space 
is very convenient to estimate their orders in $1/\N$.
Each three- or four-gluon vertex contributes a factor of $g$ or $g^2$,
respectively. Each closed index line contributes a factor of $\N$, 
while $g^2\sim 1/\N$.

\subsection{'t Hooft topological expansion}
%%\subsection{Planar diagrams}

\begin{figure}
\vspace*{3mm}
\centering{
\input{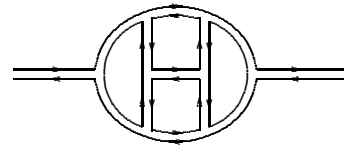}
}
\caption[]{Double-line representation of a four-loop diagram.} 
\label{fi:4loop}
\end{figure}

Let us consider a typical diagram for the gluon propagator 
depicted in Fig.~\ref{fi:4loop}.
The sum over the $\N$ indices is associated 
with each of the four closed index lines, whose number is equal
to the number of loops. The contribution of this diagram is
$\sim g^8 \N^4\sim 1$.

Diagrams of this type, which can be 
drawn on a sheet of paper without crossing any lines, are called 
{planar}\index{planar diagram} diagrams. 
For such diagrams, the addition of a loop
inevitably results in the addition of two three-gluon (or one four-gluon)
vertices. A planar diagram with $n_2$ loops has $n_2$ closed index lines.
It is of order
\be
%%\hspace*{5mm}
n_2\hbox{-loop planar diagram}~\sim~\left(g^2 \N\right)^{n_2}~\sim~1 ,
\ee
so that all planar diagrams survive in the large-$\N$ limit.   

\begin{figure}[tb]
\unitlength=.80mm
\vspace*{3mm}
\centering{
\input{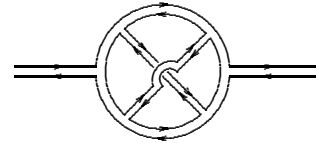}
}
\caption[]{Double-line representation of a nonplanar diagram.}   
\label{fi:genus1}
\end{figure}

Let us now consider a {nonplanar} diagram of the type depicted in 
Fig.~\ref{fi:genus1}.
The diagram has six three-gluon vertices but
only one closed index line (although it has three loops!).
The order of this diagram is $\sim g^6 \N\sim 1/\N^2$.

This nonplanar diagram  can be drawn without
line-crossing on a surface with one handle (or {hole})
which in mathematics 
is called
a torus or a surface of genus one. A plane is
then equivalent to a sphere and has genus zero. 
A general Riemann surface\index{Riemann surface} 
with $h$ holes has genus $h$. 

The above evaluations of the order of the diagrams can be described
by the single formula
\bea
\hspace*{5mm}
\hbox{genus-$h$ diagram}~\sim~
\left(\frac 1{\N^2}\right)^{{\rm genus}} .
\label{genusexpansion}
\eea
The expansion in $1/\N$ rearranges perturbation-theory diagrams
according to their topology as demonstrated in 1974 by 't~Hooft~\cite{Hoo74}. 
It is referred to 
as the {topological expansion}
\index{topological expansion!Yang--Mills theory} 
or the {genus expansion.

Only planar diagrams associated with genus zero survive
in the large-$\N$ limit. The problem of summing the 
planar graphs is complicated but simpler than that of
summing all the graphs, since the number of 
planar graphs\index{planar diagram!number of}  with
$n_0$ vertices grows geometrically at large $n_0$:
\be
%%\hspace*{5mm}
\#_{\rm p}(n_0)\;\equiv \; %%\#~
\hbox{no of planar graphs}~\sim~{\rm const}^{n_0}\,,
\label{noplanar}
\ee
as shown by Tuttle~\cite{Tut62} and Koplik, Neveu, Nussinov~\cite{KNN77},
while the total number of graphs grows factorially with $n_0$.
There is no dependence in \eq{noplanar} on the number of external
lines of a planar graph which is assumed to be much less than $n_0$.

There is a big difference between the planar diagrams
and the ladder diagrams which describe $e^+e^-$ 
elastic scattering in QED. For the ladder with $n$ rungs,  there are
$n!$ ladder diagrams, but only one of them is planar. This 
shows why the number of planar graphs is much smaller than
the total number of graphs, most of which are nonplanar.

\subsection{Topological expansion (continued)}

Equation~\rf{genusexpansion} holds, strictly speaking, only for the
gluon propagator, while the contribution of all planar diagrams to
a connected $n$-point Green function is $\sim g^{n-2}$, which is its
natural order in $1/\N$. The three-gluon Green function is $\sim g$,
the four-gluon one is $\sim g^2$ and so on.
The contributions of all planar diagrams are of the same
order $\sim 1$ in the large-$\N$ limit, independently of the number of
external lines, for the Wilson loop average 
\begin{widetext}
\be
%%\hspace*{3.5mm}
%%\lefteqn{\hspace*{-5mm}
\LA\frac1{\N}
\tr{}{\boldmath P}\e^{\i g\oint_\Gamma \d x ^\mu A_\mu \left( x \right) } \RA
= %%} \non &=& 
\sum\limits_{n=0}^\infty\; \i^n
\oint\nolimits_\Gamma \d x_1^{\mu_1}
\int\nolimits_{x_1}^{x_1} \d x_2^{\mu_2} \ldots \!
\int\nolimits_{x_1}^{x_{n-1}} \!\!\d x_n^{\mu_n}\;
G^{(n)}_{\mu_1\cdots\mu_n}\! \left(x_1,\ldots,x_n \right)
%~~~\hspace{1cm}
%%\non &&
\label{Poexpansion}
\ee
\end{widetext}
with
\be
%%\hspace*{-7mm}
G^{(n)}_{\mu_1\cdots\mu_n}\!
\left(x_1,\ldots,x_n \right) \equiv \frac {g^n}{\N} 
\LA \tr{ \!\left[A_{\mu_1}\!\left(x_1 \right) \cdots 
A_{\mu_n}\!\left(x_n \right)}\right] \,\RA.
\label{multitrnormalization}
\ee
The factor of $1/\N$, which normalizes the trace, 
provides the natural normalization.
The ordering along a closed path implies
cyclic-ordering in the index space as depicted in 
Fig.~\ref{fi:generic}, where we omit the arrows for simplicity. 
\begin{figure}[b]
\vspace*{3mm}
\centering{
\input{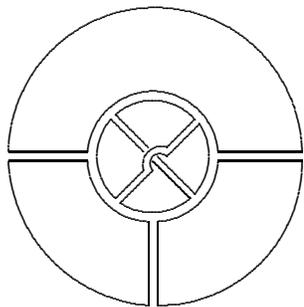}
}
\caption[]{Generic double-line index diagram.}   
\label{fi:generic}
\end{figure}
This diagram has 
 $n_0=10$ vertices, $n_1=12$ gluon
propagators, $n_2=4$ closed index lines, and $B=1$ boundary.
The color indices of the external lines are contracted 
by the Kronecker delta-symbols 
(represented by the single lines) in a cyclic order.
The extra factor of $1/\N$ arises from the normalization.
The order in $1/\N$ 
of the diagram in Fig.~\ref{fi:generic} is $\sim 1/\N^2$ 
in accord with \eq{genusexpansion}.

Analogously,
the color indices in \eq{multitrnormalization} are contracted in 
the cyclic order.
The delta-symbols, which contract the color indices, are
depicted by the single lines. They can be viewed as a {boundary} of 
the  diagram. The actual size of the boundary is not essential --
it can be shrunk to a point. Then a bounded piece of a plane will be
topologically equivalent to a sphere with a puncture.
We draw planar diagrams in a plane with an extended
boundary (boundaries) rather than in a sphere with a puncture (punctures). 
The closed boundary is associated with the trace over the color 
indices of the multi-point Green function

The boundary represents the Wilson loop = a trajectory of
a heavy quark in the fundamental representation.

\subsection{Topological expansion and quark loops\label{ss:t.e.}}

It is easy to incorporate quarks in 
the topological expansion.
\index{topological expansion!Yang--Mills theory!with quarks}
A quark field belongs to the fundamental representation of the 
gauge group $SU(\N)$ and its propagator is represented by 
a single line
\be
\left\langle  \psi _i\bar\psi _j\right\rangle ~ \propto ~ \delta _{ij} ~=
\hspace*{-4mm}
\unitlength=1mm
\linethickness{0.6pt}
\begin{picture}(31.00,4.00)(10,10)
%\begin{picture}(21.00,12.00)(0,6)
\put(20.00,11.00){\vector(1,0){6.00}}
\put(26.00,11.00){\line(1,0){4.00}}
\put(18.00,11.00){\makebox(0,0)[cc]{$i$}}
\put(32.00,11.00){\makebox(0,0)[cc]{$j$}}
\end{picture}
\hspace*{-7mm}.
\label{singleline}
\ee
The arrow indicates, as usual, the direction of propagation of
a (complex) field $\psi$. These arrows 
are often omitted for simplicity.

The diagram for the gluon propagator which involves one quark loop
is depicted in Fig.~\ref{fi:quarkl}a.
\begin{figure}
\vspace*{3mm}
\centering{
\input{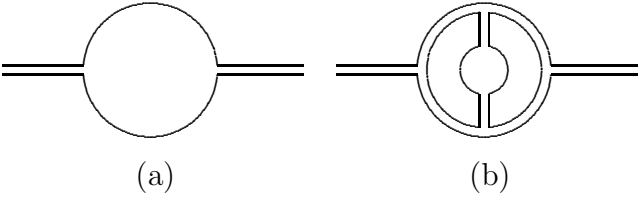}
}
\caption[]{Diagrams for gluon propagator which involve quark loop.}   
\label{fi:quarkl}
\end{figure}
It involves one quark loop and has no closed index lines 
so that its order is $\sim g^2\sim 1/\N$. The diagram in Fig.~\ref{fi:quarkl}b
is $\sim g^6 \N^2 \sim 1/\N$ analogously.

It is evident from this consideration that 
quark loops are not\index{quark loop at large $\N$}
accompanied by closed index lines.\index{index line} 
One should add a closed index line
for each quark loop in order for a given diagram with $L$ quark
loops to have the same 
double-line representation  %%\index{double-line representation} 
as for pure gluon diagrams.
Therefore, given \eq{genusexpansion}, diagrams with $L$ 
quark loops\index{quark loop at large $\N$}
are suppressed at large $\N$ by 
\bea
\hspace*{5mm}
L~\hbox{quark loops}~\sim~\left(\frac1{\N}\right)^{L+2\cdot{\rm genus}}.
\label{Lqloops}
\eea

The single-line representation of 
the quark loops\index{quark loop at large $\N$} 
is similar to that of 
the Wilson loop. Such a diagram
emerges in gluon corrections to the
vacuum expectation value of the quark operator
\bea
\hspace*{5mm}
O~=~\frac 1{\N} \bar \psi \psi \,,
\label{QO}
\eea
where the factor of $1/\N$ is introduced to make it ${\cal O}(1)$ in 
the large-$\N$ limit. Therefore, the external 
boundary can be viewed as a single line associated with 
valence quarks.\index{valence quark}

\subsection{The proof of topological expansion}

To prove Eqs.~\rf{genusexpansion} and its quark 
counterpart~\rf{Lqloops},
let us consider a generic diagram in the index space which has
$n_0^{(3)}$ three-point vertices (either three-gluon or quark--gluon ones),
$n_0^{(4)}$ four-gluon vertices, $n_1$ propagators 
(either gluon or quark ones), 
$n_2$ closed index lines, $L$ virtual quark loops and $B$ external boundaries. 
Its order in $1/\N$ is 
\bea
\frac1{\N^B}g^{n_0^{(3)}+2 n_0^{(4)}} \N^{n_2}&\sim&
\N^{n_2 - n_0^{(3)}\!/2 - n_0^{(4)}-B}
\label{ordergeneric}
\eea
as has already been explained. The extra factor of $1/\N^B$ arises from 
the extra normalization factor of $1/\N$ in operators associated with
external boundaries.

The number of propagators and vertices
are related by
\bea
\hspace*{5mm}
2 n_1~=~3 n_0^{(3)}+ 4 n_0^{(4)} ,
\label{234}
\eea
since three- and four-point vertices emit three or four propagators,
respectively, and each propagator connects two vertices.
We then rewrite the RHS of~\rf{ordergeneric} as
\bea
\N^{n_2 - n_0^{(3)}\!/2 - n_0^{(4)}-B}&=&\N^{n_2-n_1+n_0-B} ,
\label{ordergenerice}
\eea
where $n_0=n_0^{(3)}+ n_0^{(4)}$ is the total number of vertices.

The exponent on the RHS of \eq{ordergenerice} can be expressed via the 
Euler characteristic\index{Euler characteristic} 
$\chi$ of a given graph of genus $h$.
An appropriate 
Riemann surface\index{Riemann surface}, which is
associated with a given graph, is open and has $B+L$ boundaries.
This surface can be closed by attaching
a cap to each boundary. The single lines then become double lines
%%\index{double-line representation}
together with the lines of the boundary of each cap. We have already 
considered this procedure when deducing \eq{Lqloops} from \eq{genusexpansion}.

The number of faces for a
closed Riemann surface\index{Riemann surface} 
constructed in such a manner is \mbox{$n_2+L+B$}, 
while the number of edges and vertices are $n_1$ and $n_0$, respectively.
Euler's theorem\index{Euler characteristic} states that
\bea
\chi~\equiv~2-2h&=&n_2+L+B-n_1+n_0\,.
\label{Euler}
\eea
Therefore the RHS of \eq{ordergenerice} can be rewritten as
\bea
\N^{n_2-n_1+n_0-B}&=&\N^{2-2h-L-2B} .
\eea

We have thus proven that the order in $1/\N$ of a generic graph
does not depend on its order in the coupling constant and is 
completely expressed via the genus $h$ and the number 
of virtual\index{virtual quark} quark loops\index{quark loop at large $\N$} 
$L$ and external boundaries $B$ by
\bea
\hspace*{5mm}
\hbox{generic graph}~\sim~\left(\frac1{\N}\right)^{2h+L+2(B-1)} .
\label{ordergenericee}
\eea 
For $B=1$, we recover Eqs.~\rf{genusexpansion} and \rf{Lqloops}.
 
\subsection{'t~Hooft versus Veneziano limits\label{ss:tHvsV}}

\index{'t~Hooft limit}\index{Veneziano limit}
In QCD there are several species or flavors\index{flavor} of quarks 
($u$-, $d$-, $s$- and so on). We denote the number of flavors\index{flavor}
by $\Nf$ and associate a Greek letter $\alpha$ or $\beta$ with
a flavor\index{flavor} index of the quark field.

The quark propagator then has the Kronecker delta-symbol with 
respect to the flavor\index{flavor} indices in addition to \eq{singleline}:
\begin{eqnarray}
\left\langle \psi _i^\alpha \bar \psi _j^\beta \right\rangle ~&
\propto & \delta^{\alpha \beta }\delta _{ij} \,.
\end{eqnarray}
Their contraction results in
\bea
\sum_{\alpha=1}^{\Nf} \delta_{\alpha\alpha}&=&\Nf\,.
\label{N_f}
\eea
Therefore, an extra factor of $\Nf$ corresponds to each closed quark
\index{quark loop at large $\N$}
loop for the $\Nf$ flavors.\index{flavor}

The limit when $\Nf$ is fixed as $\N\to\infty$ is called 
the {'t~Hooft limit}.\index{'t~Hooft limit}\/
Only valence quarks\index{valence quark} 
are then left
 (the quenched approximation). In order for a meson
to decay into other mesons built out of quarks, a quark--antiquark pair must
be produced out of the vacuum. Consequently, the ratios of meson widths to
their masses are
\bea
\frac{\Gamma_{\rm total}}{M}&\sim&\frac{\Nf}{\N} 
\label{width}
\eea
in the 't~Hooft limit.\index{'t~Hooft limit} 
The ratio on the LHS of \eq{width} is 
$10$--$15$\% experimentally for the $\rho$-meson. 
The hope of solving QCD in the 't~Hooft
limit\index{'t~Hooft limit} 
is the hope to describe QCD with this accuracy.

An alternative large-$\N$ limit of QCD, when $\Nf\sim \N$ as $\N\to\infty$,
was proposed by Veneziano in 1976~\cite{Ven76}.
A general diagram with 
$L$ quark loops\index{quark loop at large $\N$} will contribute
\bea
\hspace*{5mm}
L~\hbox{quark loops}~ \sim~\left(\frac{\Nf}{\N} \right)^L
\left(\frac{1}{\N^2} \right)^{\rm genus},
\label{fLqloops}
\eea 
since each quark loop results in $\Nf$.

The quark loops\index{quark loop at large $\N$} 
are not suppressed at large $\N$ in the 
{Veneziano limit}\index{Veneziano limit} 
\begin{eqnarray}
\hspace*{5mm}
\Nf~\sim \N~ \rightarrow ~\infty 
\label{V.l.}
\end{eqnarray}
if the diagram is planar.

It is the Veneziano limit  
that is related to the hadronic
topological expansion in the dual-resonance models. In the Veneziano limit 
hadrons can have
finite widths according to \eq{width}.

\section{Large-$\N$ factorization\label{ss:mfac}} 

\index{factorization at large $N$!Yang--Mills theory}
The vacuum expectation values of several colorless or white operators,
which are singlets with respect to the gauge group, factorize in
the large-$\N$ limit of QCD (or other matrix models) as was
first noticed by A.~A.~Migdal and independently E.~Witten in late 1970's. 

The simplest gauge-invariant operators in a pure $SU(\N)$ gauge theory
are the closed Wilson loops
\be
\Phi( C)=
\frac {1}{\N}\,{\rm tr}\,P \e^{\i g \oint_C \d z^\mu A_\mu(z)}\,.
\label{defO}
\ee
They obey the factorization property
\be
\LA \,\Phi(C_1)\cdots \Phi(C_n)\,\RA=\LA\, \Phi(C_1) \,\RA \cdots 
\LA\, \Phi(C_n) \,\RA
+{\cal O}\!\left(\N^{-2}\right) .
\label{mfactorization}
\ee

The factorization implies a semiclassical nature of the
large-$\N$ limit of QCD (a saddle point in the path integral
for certain variables).

The factorization property
\index{factorization at large $N$!Yang--Mills theory}  
also holds for gauge-invariant
operators constructed from quarks as in \eq{QO}. For the case of
several flavors\index{flavor} $\Nf$, we normalize these 
quark operators\index{quark operator} by
\be
O_\Gamma=\frac 1{\Nf \N} \bar \psi \Gamma \psi \,.
\label{OGamma}
\ee
Here $\Gamma$ denotes one of the combination of 
the $\gamma$-matrices:\index{matrices!$\gamma$}
\be
\Gamma~=~{\boldmath I},\;\gamma_5,\;\gamma_\mu,\;\i\gamma_\mu \gamma_5,\;
\Sigma_{\mu\nu}=\frac{1}{2\i}[\gamma_\mu,\gamma_\nu]\;,\ldots\;.
\ee

The factorization
\index{factorization at large $N$!Yang--Mills theory!quark operator}  
of the gauge-invariant 
quark operators\index{quark operator} 
holds both in the 't~Hooft\index{'t~Hooft limit} 
and Veneziano limits:\index{Veneziano limit}
\be
\LA \,O_{\Gamma_1}\cdots 
O_{\Gamma_n}\,\RA=\LA\, O_{\Gamma_1} \,\RA \cdots \LA\, O_{\Gamma_n} \,\RA
+{\cal O}\!\left(1/ (\Nf\N) \right).
\label{QQQQ}
\ee
The nonfactorized part, which is associated with connected diagrams,
is $\sim 1/\N$ in the 't~Hooft limit.\index{'t~Hooft limit} 
This leads, in particular, 
to the coupling constant of meson--meson interaction of order 
$1/\N$.
The Veneziano limit is analogous to pure Yang-Mills.\index{Veneziano limit} 

The factorization
\index{factorization at large $N$!Yang--Mills theory!quark operator}   
can be seen  (at all orders of perturbation theory) 
from \eq{ordergenericee} for the contribution of a generic
connected graph of genus $h$ with $B$ external boundaries which are
precisely associated with the Wilson loops (or
the quark operators\index{quark operator} 
$O_{\Gamma}$). The diagrams with gluon lines emitted and
absorbed by the same operator are
products of diagrams having only one boundary.
Their contribution is of order one. The diagrams
with gluon lines emitted and absorbed by two different operators 
have two boundaries.
This proves the factorization
\index{factorization at large $N$!Yang--Mills theory!quark operator} 
property~\rf{QQQQ} at all orders of perturbation theory.

The large-$\N$ factorization
\index{factorization at large $N$!Yang--Mills theory}  
can also be verified beyond perturbation theory
at all
orders of the strong-coupling expansion
\index{strong-coupling expansion!factorization in} 
in the $SU(\N)$ lattice
gauge theory. A nonperturbative proof 
of the factorization\index{factorization at large $N$!Yang--Mills theory}  
was given using quantum equations of motion (the loop equations)~\cite{MM79}. 

\section{Conclusion}

We have considered in these lecture notes the basic features of the methods
for nonperturbative studies of gauge theories, which were developed
in the second half of 1970's -- early 1980's. Their contemporary applications
in high-energy physics are extremely broad: from the scattering of particles
at very high energies to the attempts of constructing a unified theory of
all interactions, including gravity. Some of these issues are considered
in other lectures at this School.

%%\end{document}

%%\eop

\end{document}